\documentclass[useAMS,usenatbib,twocolumn]{mn2e}
\usepackage[dvips]{graphicx}
\usepackage{amsmath}
\usepackage{amsfonts}
\usepackage{amssymb}

\title[Mulitbody Asterodensity Profiling (MAP)]
{A Novel Method to Photometrically Constrain Orbital Eccentricities: 
Multibody Asterodensity Profiling (MAP)}
\author[Kipping et al.]{David M. Kipping$^{1,2}$\thanks{E-mail: dkipping@cfa.harvard.edu}, William R. Dunn$^{3,\dagger}$, Jamie M. Jasinski$^{3,\dagger}$ \& Varun P. Manthri$^{3,\dagger}$ \\
$^{1}$Harvard-Smithsonian Center for Astrophysics, 60, Garden St., Cambridge, MA 02138, USA\\
$^{2}$Carl Sagan Fellow\\
$^{3}$Department of Physics and Astronomy, University College London, Gower St., London WC1E 6BT\\
$^{\dagger}$These authors contributed equally to the work presented here}
\begin{document}

\date{Accepted 2011 December 12. Received 2011 December 12; in original form 2011 October 23}

\pagerange{\pageref{firstpage}--\pageref{lastpage}} \pubyear{2011}

\maketitle

\label{firstpage}


\begin{abstract}

We present a novel method to determine eccentricity constraints of extrasolar
planets in systems with multiple transiting planets through photometry alone. 
Our method is highly model independent, making no assumptions about the stellar 
parameters and requiring no radial velocity, transit timing or occultation 
events. Our technique exploits the fact the light curve derived stellar density 
must be the same for all planets transiting a common star. Assuming a circular 
orbit, the derived stellar density departs from the true value by a predictable
factor, $\Psi$, which contains information on the eccentricity of the system. By 
comparing multiple stellar densities, any differences must be due to 
eccentricity and thus meaningful constraints can be placed in the absence of any 
other information. The technique, dubbed ``Multibody Asterodensity Profiling'' 
(MAP), is a new observable which can be used alone or in combination with
other observables, such as radial velocities and transit timing variations.
An eccentricity prior can also be included as desired. MAP is most
sensitive to the minimum pair-combined eccentricity e.g. 
$(e_1 + e_2)_{\mathrm{min}}$. Individual eccentricity constraints are less
stringent but an empirical eccentricity posterior is always derivable and
is freely available from transit photometry alone.

We present a description of our technique using both analytic and numerical 
implementations, followed by two example analyses on synthetic photometry as a 
proof of principle. We point out that MAP has the potential to constrain the 
eccentricity, and thus habitability, of Earth-like planets in the absence of 
radial velocity data, which is likely for terrestrial-mass objects.

\end{abstract}



\section{Introduction}
\label{sec:intro}

In February of 2011, 1235 \emph{Kepler} transiting candidate planets were
announced by \citet{borucki:2011}, amongst which the majority are expected to be 
genuine \citep{morton:2011}. At the latest counting, the score has since risen
to 1781 \citep{rowe:2011} and is expected to continue rising. Due to the 
unprecedented yield of new transiting planet candidates, follow-up with radial 
velocity (RV) measurements is generally not feasible due to both the typical 
faintness of the targets and the intensive nature of the required telescope time 
for so many targets. Historically, radial velocity has emerged as the tool of 
choice to confirm transiting candidates and so the \emph{Kepler} team have 
devoted considerable effort to find ways to confirm candidates without the need 
for RV. This has led to some pioneering techniques such as blend analysis 
\citep{torres:2011,fressin:2011} and confirmation through transit timing 
variations (TTV) \citep{holman:2011,lissauer:2011}.

However, even though transiting candidates have been shown to be confirmable
without RV, its absence means that the orbital eccentricity, $e$, of the 
planets cannot be determined (unless very strong TTVs are detected). One 
remaining avenue to constrain $e$ is to detect an occultation event. 
Occultations occur exactly half an orbital period after the transit event for a 
circular orbit\footnote{There also exists a small light travel time across the 
system} but become offset for eccentric orbits, thus offering a potential 
diagnostic of eccentricity (details on the precise obtainable constraints are 
provided in \citealt{thesis:2011}). Such occultations are due to a combination 
of reflected light from the planet and thermal emission (which is very small in 
\emph{Kepler}'s visible bandpass e.g. \citealt{darkest:2011}). Despite 
\emph{Kepler's} ground-breaking photometric precision, most transiting planets 
discovered so far have not exhibited such events (one notable exception is the 
high albedo planet Kepler-7b, see \citealt{kippingbakos:2011a}). Therefore, in 
the majority of cases we are left without any way of characterizing the orbital 
eccentricity.

Our principal motivation for addressing this problem stems from the fact that 
based upon the estimated frequency of Earth-like planets 
\citep{howard:2010,catanzarite:2011,wittenmyer:2011}, it seems probable that 
\emph{Kepler} will detect numerous habitable-zone Earth-radius planets. After
the initial detection, the natural question we will be ``can this world sustain 
life?''. A high orbital eccentricity causes a planet to spend the majority of 
its orbit outside the habitable-zone and thus leads to potentially marginal or 
transient habitability \citep{williams:2002,dressing:2010}. For example, 
eccentricities in excess of $0.3$ for a habitable-zone planet around a Sun-like 
star cause equilibrium temperatures to vary by more than 100\,K. The ability to 
measure, or at least constrain, eccentricity therefore would greatly benefit the 
assessment of an exoplanet's habitability.

Recently, \citet{moorhead:2011} (M11 from here on in) have studied the 
problem in an effort to glean some information about the eccentricity of a 
transiting planet. Their approach is that if one knows the stellar density 
a-priori, say from stellar spectroscopy combined with evolution models, then one 
can compute the maximum allowed transit duration for a planet on a circular 
orbit using (note that this approach was originally outlined in 
\citealt{ford:2008}):

\begin{equation}
\tilde{T}_{\mathrm{max}}^{\mathrm{circ}} \simeq \frac{R_* P}{\pi a}
\end{equation}

where $\tilde{T}$ is the duration between the planet's centre crossing the
stellar limb to exiting under the same condition, $R_*$ is the stellar radius,
$P$ is the planet's orbital period and $a$ is the planet's semi-major axis. The
above equation assumes an equatorial transit and hence is the maximum duration
possible. M11 discuss how if the observed transit duration
exceeds this quantity (i.e. 
$\tilde{T}/\tilde{T}_{\mathrm{max}}^{\mathrm{circ}}>1$), this indicates that
the orbit must be eccentric. The two weaknesses in this approach are that:
1) The spectroscopically determined value of $R_*$ has both a large statistical
uncertainty and a large and unknown systematic uncertainty (for example,
\citet{brown:2011} state that the Kepler Input Catalogue effective temperatures 
and radii estimations are reliable for Sun-like stars, but are ``untrustworthy'' 
for stars with $T_{\mathrm{eff}}<3750$\,K).
2) A planet can be eccentric yet still cause 
$\tilde{T}/\tilde{T}_{\mathrm{max}}^{\mathrm{circ}}<1$ i.e. only planets
transiting near to apoapse, the slowest part of the orbit, will be identified
as eccentric, which is for $\omega\sim270^{\circ}$. The first weakness means 
the technique is model-dependent and that an individual system may not be 
reliable due to possible systematic errors in $R_*$. The second weakness limits 
the scope of application of the technique to planets with $b\sim0$ and
$\omega\sim270^{\circ}$, which it should be noted is the least probable value of 
$\omega$ from geometric priors \citep{kane:2008}.

M11 show that the $R_*$ uncertainty weakness can be overcome
by adopting a statistical perspective. Even though an individual system may not
be reliable, the bulk of systems should be and so any overall distributions 
which emerge should be reliable. The valuable technique of M11
allows us to actually say something about the eccentricities of the 
\emph{Kepler} candidates as a whole.

But what about individual systems? Or those which don't both fortuitously
transit near $\omega\sim270^{\circ}$ and have $b\sim0$? If the star has been 
studied with asteroseismology, then this prior can solve the riddle (as pointed 
out in M11). This is scenario is discussed in more detail
later in \S~\ref{sub:SAP}. However, to date, relatively few targets have had 
asteroseismology studies conducted. We are therefore left with the quandary that 
it is usually impossible to say anything empirical about the orbital 
eccentricity of a transiting planet through photometry alone\footnote{Unless an 
occultation is observed}.


\section{MAP: Analytic Constraints (\MakeLowercase{a}-MAP)}
\label{sec:a-MAP}

\subsection{Multiple Transiting Planet Systems}
\label{sub:multis}

We here present a solution for solving this riddle, which is applicable for 
multiple transiting planet systems. Although this may seem limited in 
application, in fact nearly 50\% of all transiting planet candidates discovered 
by \emph{Kepler} are in multi-planet systems \citep{rowe:2011}.
Table~\ref{tab:multis} provides the most recent statistics on multiple-planet
systems from \emph{Kepler}. With this point established, we will now outline
our proposed method.

\begin{table*}
\caption{\emph{Statistics regarding the number of planetary candidates
found in multiple systems by \emph{Kepler}. In this table, all examples of
the word ``planet'' refer to a planetary candidate. Also, a ``multi-system''
refers to a solar system containing more than one transiting planet candidate.}} 
\centering 
\begin{tabular}{c c c} 
\hline
Statistic & As of \citet{borucki:2011} & As of \citet{rowe:2011} \\ [0.5ex] 
\hline
Number of planets & 1235 & 1781 \\
Number of systems with planets & 997 & 1296 \\
Number of multi-systems & 170 & 328 \\
Number of single-systems & 827 & 968 \\ 
\% of systems which are multi-systems & 17.1\% & 25.3\% \\
\% of systems which are single-systems & 82.9\% & 74.7\% \\
Number of planets in a multi-system & 408 & 813 \\
Number of planets in a single-system & 827 & 968 \\
\% of planets in a multi-system & 33.0\% & 45.6\% \\ 
\% of planets in a single-system & 67.0\% & 54.4\% \\ 
2 planet systems & 115 & 218 \\
3 planet systems & 45 & 75 \\
4 planet systems & 8 & 25 \\
5 planet systems & 1 & 8 \\
6 planet systems & 1 & 2 \\ [1ex]
\hline\hline 
\end{tabular}
\label{tab:multis} 
\end{table*}

\subsection{Light Curve Derived Stellar Density}
\label{sub:rhocirc}

When a planet transits a star, consecutive transits provide the orbital period, 
$P$, and the light curve morphology contains information about the semi-major 
axis scaled in units of the stellar radius, $a/R_*$ 
\citep{seager:2003,investigations:2010}. If it were possible to get $a$ 
directly, then we use Kepler's Third Law to determine $M_*$ (we here assume 
$M_P\ll M_*$):

\begin{align}
M_* &= \frac{4\pi^2 a^3}{G P^2}
\end{align}

Unfortunately, this is not possible. The light curve only lets us measure
$a/R_*$. Adding this into the above equation means we get the following:

\begin{align}
\frac{M_*}{R_*^3} &= \frac{4\pi^2 a^3}{G P^2 R_*^3} \nonumber \\
\rho_* &= \frac{3\pi (a/R_*)^3}{G P^2}
\label{eqn:rhostar}
\end{align}

Therefore, we can determine the stellar density from the transit light curve.
This well-known trick, first pointed out by \citet{seager:2003}, has been a
powerful instrument in the toolbox of the exoplanetary scientist. It is common
practice to determine $\rho_*$ from the light curve and then use stellar
evolution models to estimate $M_*$ and $R_*$ separately.

\subsection{Eccentric Orbits}
\label{sub:rhoecc}

As we discussed earlier, without RV or an occultation we have no way of knowing 
what the orbital eccentricity of a transiting planet is. The major problem with
this is that the determined value of $a/R_*$ is heavily affected by orbital
eccentricity. If we assume a circular orbit, but the orbit is really eccentric,
then the determined value of $a/R_*$ would erroneously be (see 
\citealt{investigations:2010} for proof):

\begin{align}
(a/R_*)^{\mathrm{circ}} &\simeq \frac{1+e\sin\omega}{\sqrt{1-e^2}} (a/R_*)
\end{align}

In other words, $a/R_*$ is wrong by a factor given by 
$(1+e\sin\omega)/\sqrt{1-e^2}$. Note that the above equation is an approximate 
formula based upon the $\tilde{T}^{\mathrm{one}}$ expression for the transit
duration \citep{investigations:2010}. The reliability of this approximate
expression will be dealt with later in \S\ref{sub:psireliability}.

If we determine a biased value for $a/R_*$, then one will determine a biased 
value for $\rho_*$ too, since $\rho_*$ depends upon $a/R_*$, as seen in 
Equation~\ref{eqn:rhostar}. This means that the derived stellar density becomes

\begin{align}
\rho_{*}^{\mathrm{circ}} &\simeq \rho_* \Psi
\label{eqn:psieqn}
\end{align}

where

\begin{align}
\Psi &\equiv \frac{(1+e\sin\omega)^3}{(1-e^2)^{3/2}}.
\end{align}

\subsection{Double Transiting Systems}
\label{sub:doubles}

Consider two planets, dubbed with subscripts ``1'' and ``2'', which
have been observed to transit the same star. Let us assume that we fit these
light curves assuming $e_1=0$ and $e_2=0$, since we have no RVs, occultation or
strong TTVs and hence no reason to assume otherwise. For each planet, one may
empirically determine $\rho_{*,k}^{\mathrm{circ}}$:

\begin{align}
\rho_{*,1}^{\mathrm{circ}} &= \Psi_1 \rho_* \nonumber \\
\rho_{*,2}^{\mathrm{circ}} &= \Psi_2 \rho_*
\end{align}

Since the parent star is the same for planets, then one can divide these two 
equations to give:

\begin{align}
\Bigg(\frac{\rho_{*,1}^{\mathrm{circ}}}{\rho_{*,2}^{\mathrm{circ}}}\Bigg) &= \frac{\Psi_1}{\Psi_2} \nonumber \\
\Bigg(\frac{\rho_{*,1}^{\mathrm{circ}}}{\rho_{*,2}^{\mathrm{circ}}}\Bigg) &= \Bigg(\frac{1+e_1 \sin \omega_1}{1+e_2 \sin \omega_2}\Bigg)^3 \Bigg(\frac{1-e_2^2}{1-e_1^2}\Bigg)^{3/2}
\label{eqn:rhoratio}
\end{align}

Note, that a similar equation to this appears in \citet{ragozzine:2010},
Equation~3, although the equation is known to contain an error (D. Ragozzine 
personal communication). Nevertheless, the authors hint at the possibility
that it may possible to determine eccentricities via this ratio.

We stress here that the term on the left-hand-side of 
Equation~\ref{eqn:rhoratio} is an observable. One can therefore see that it is 
possible to glean some information about the eccentricity of the system. 
However, the current form is not very informative. We have one observable, 
$(\rho_{*,1}^{\mathrm{circ}}/\rho_{*,2}^{\mathrm{circ}})$, and four 
unknowns: $e_1$, $\sin\omega_1$, $e_2$ and $\sin\omega_2$.

Ultimately, the quantity which will affect potential habitability of a planet
will be $e_k$ and not $\omega_k$. Therefore, the $\omega_k$ quantity is of lower
interest to us. The $\sin\omega_k$ terms must lie in the range 
$-1\leq\sin\omega_k\leq+1$, which means we can construct a lower and upper bound
inequality for the quantity $\Theta_{12}$:

\begin{align}
\Theta_{12} &= \Bigg(\frac{\rho_{*,1}^{\mathrm{circ}}}{\rho_{*,2}^{\mathrm{circ}}}\Bigg)^{2/3} \nonumber \\
\Theta_{12} &= \Bigg(\frac{1+e_1 \sin \omega_1}{1+e_2 \sin \omega_2}\Bigg)^2 \Bigg(\frac{1-e_2^2}{1-e_1^2}\Bigg)
\label{eqn:theta}
\end{align}

Which must satisfy:

\begin{align}
\Bigg(\frac{1-e_1}{1+e_1}\Bigg) \Bigg(\frac{1-e_2}{1+e_2}\Bigg) \leq \Theta_{12} \leq \Bigg(\frac{1+e_1}{1-e_1}\Bigg) \Bigg(\frac{1+e_2}{1-e_2}\Bigg)
\label{eqn:eccinequalityA}
\end{align}

One may now replace $e_2 = \epsilon_{21} e_1$ where $\epsilon_{21} = e_2/e_1$
and expand to first-order in $e_1$:

\begin{align}
-(e_1+e_2) + \mathcal{O}[e_1]^2 &\leq \frac{\Theta_{12}-1}{2} \leq (e_1+e_2) + \mathcal{O}[e_1]^2 \nonumber \\
\therefore e_1+e_2 &\gtrsim \frac{\Theta_{12}-1}{2}
\label{eqn:doubleA}
\end{align}

If one instead replaces $e_1 = \epsilon_{12} e_2$ in 
Equation~\ref{eqn:eccinequalityA} and expands to first order in $e_2$, an 
identical equation is obtained.

Before moving onto triple systems, we discuss a final subtlety. The fraction 
$\Theta_{12}$ can be inverted to give $\Theta_{21}$ and yet the
same inequality is derivable as Equation~\ref{eqn:eccinequalityA}, namely we
have:

\begin{align}
\Bigg(\frac{1-e_1}{1+e_1}\Bigg) \Bigg(\frac{1-e_2}{1+e_2}\Bigg) \leq \Theta_{21} \leq \Bigg(\frac{1+e_1}{1-e_1}\Bigg) \Bigg(\frac{1+e_2}{1-e_2}\Bigg)
\label{eqn:eccinequalityB}
\end{align}

Following the same steps as before, means we arrive at:

\begin{align}
e_1+e_2 &\gtrsim \frac{\Theta_{21}-1}{2}
\label{eqn:doubleB}
\end{align}

So we have two equations, both of which must hold, for constraining the
$(e_1+e_2)$ combination. Both Equation~\ref{eqn:doubleA}\&\ref{eqn:doubleB}
require that $\Theta_{ij}>1$ in order to place positive constraints on this
value (a negative value has no physical meaning). Clearly, if $\Theta_{12}<1$
then $\Theta_{21}>1$ and vice versa. Thus we can always construct a
physically meaningful constraint on $(e_1+e_2)$ by taking the maximum of the 
two.

In practice, we wish to produce a posterior distribution of $(e_1+e_2)$ based
upon the posterior of $\Theta_{12}$ or $\Theta_{21}$. We can choose to use
either version of $\Theta_{ij}$ but not both i.e. we cannot create a posterior
which swaps between the two $\Theta_{ij}$ versions. The simplest thing is
to produce two posteriors and then select the one which provides the
most meaningful constraints. This selection can be done visually, or by
say taking the median of both posteriors and choosing the largest.

\subsection{Reliability of the a-MAP Inequality}
\label{sub:aMAPreliability}

\begin{figure}
\begin{center}
\includegraphics[width=8.4 cm]{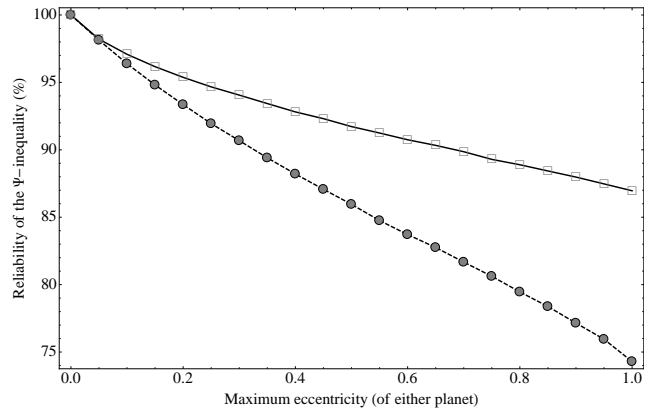}
\caption{\emph{The reliability of the inequality in Equation~\ref{eqn:doubleA}
as a function of the maximum allowed eccentricity, assuming a uniform 
distribution in $e_k$ (dashed line) and also a non-uniform physically motivated 
distribution (solid line). The reliability is $\geq$90\% for 
$e_k\leq0.30$ for the uniform case, and $e_k\leq0.65$ for the non-uniform 
case.}} 
\label{fig:psis}
\end{center}
\end{figure}

Equations~\ref{eqn:doubleA}\&\ref{eqn:doubleB} are approximate equations
valid to first order in $e_k$ only. Therefore, the reliability of the inequality
will deteriorate for large $e_k$. To test the accuracy of these expressions, we 
generated some random values for $e_1$, $\omega_1$, $e_2$ and $\omega_2$. The 
$\omega_k$ values have uniform distributions between 0 and $2\pi$ and the $e_k$ 
values have uniform distributions between $0$ and $e_{\mathrm{max}}$. We 
generated these random values $10^6$ times and tested if the inequality in 
Equation~\ref{eqn:doubleA} was true or not each time (the accuracy of 
Equation~\ref{eqn:doubleA} will be the same as that of 
Equation~\ref{eqn:doubleB} due to symmetry arguments). As an example, using 
$e_{\mathrm{max}} = 0.25$, the inequality is true in 91.9\% of all of the Monte 
Carlo simulations. In Figure~\ref{fig:psis}, we show the percentage of trials 
for which the inequality is correct as a function of $e_{\mathrm{max}}$, which 
reveals that the inequality provides useful eccentricity constraints in the 
absence of any other information and is $\geq$90\% reliable for 
$e_{\mathrm{max}} \leq 0.30$. 

We also tried using a potentially more realistic non-uniform distribution 
eccentricity distribution using a mixture of an exponential and a Rayleigh
distribution (see \citealt{juric:2008}; \citealt{zakamska:2011}):

\begin{align}
\mathrm{P}(e_k) &= \alpha \lambda \exp[-\lambda e_k] + (1-\alpha) \frac{e_k}{\sigma_e^2} \exp[-e_k^2/(2\sigma_e^2]
\label{eqn:eprior}
\end{align}

The values of the constants were found by fitting the distribution of 
eccentricities in known multi-planet systems measured from radial velocity 
surveys using only systems with measured eccentricities, which find
$\alpha =0.38$, $\lambda=15$ and $\sigma_e = 0.17$ \citep{steffen:2010}. 
Finally, this distribution can produce values of $e_k$ greater than unity, and 
so we ignored any simulations where $e_P > e_{\mathrm{max}}$ for either planet. 
Using $e_{\mathrm{max}} = 1$, we found that 87.0\% of simulations agreed with 
the inequality presented in Equation~\ref{eqn:doubleA}, and $\geq$90\% agree for 
$e_{\mathrm{max}}\leq0.65$ (Figure~\ref{fig:psis} shows dependency of this 
percentage with $e_{\mathrm{max}}$).

\subsection{Triple Transiting Systems}
\label{sub:triples}

For three-planet systems, one may construct three $\Theta_{ij}$ ratios meaning
we now have six unknowns and three observables. By blanketing out the $\omega_k$
terms in a similar way described for double-transiting systems (and thus
limiting ourselves to lower bounds on $e_k$) we are left with three unknowns
($e_1$, $e_2$ \& $e_3$) and three observables ($\Theta_{12}$, $\Theta_{23}$ \&
$\Theta_{31}$, following a numerical cyclic). One would therefore presume that 
it should be possible to constrain the eccentricities individually rather than 
limiting ourselves to combination terms. In total, we have three inequalities, 
in analogy to the double-transiting case:

\begin{align}
e_1+e_2 &\gtrsim \frac{1}{2} \{ \Theta_{12}-1 ,  \Theta_{21}-1\} \\
e_2+e_3 &\gtrsim \frac{1}{2} \{ \Theta_{23}-1 ,  \Theta_{32}-1\} \\
e_3+e_1 &\gtrsim \frac{1}{2} \{ \Theta_{31}-1 ,  \Theta_{13}-1\}
\label{eqn:triple}
\end{align}

One may naively assume that these may converted into individual limits by 
solving simultaneously. However, such an operation requires subtracting one
inequality from another, which is a strictly forbidden operation. With this
bar, it is not possible to solve the expressions and so the furthest we can
ever take a-MAP is to produce the inequalities of 
Equations~\ref{eqn:doubleA}\&\ref{eqn:doubleB}. Naturally, this leads us to
consider alternative methods which are not based upon analytic methods. However,
before we do, we will pause to evaluate the range of parameters under which
Equation~\ref{eqn:psieqn} (the $\Psi$-equation) is valid. This is crucial
since all of the a-MAP expressions are built around this equation and even the
later numerical techniques have the same dependency.

\subsection{Reliable Parameter Range for the $\Psi$-Equation}
\label{sub:psireliability}

Equation~\ref{eqn:psieqn} has been revealed to be the key to unlocking
some information about the eccentricity of transiting planets. It is worth,
though, pausing to evaluate the reliability of this expression, since the
equation is an approximation, as explicitly stated in 
\citet{investigations:2010}. It should also be stressed that the non-analytic 
method discussed later (\S\ref{sec:n-MAP}) relies on the simplicity of the 
$\Psi$-equation too, and would not work using a more elaborate form. 
The reliable range of the $\Psi$ equation informs the reliable
range of our proposed technique in general, and thus is crucial for 
understanding the range of applicability of our new method.

The $\Psi$-equation describes how erroneous the derived stellar density would be
if we assumed a circular orbit for an eccentric planet. As was seen in
Equation~\ref{eqn:rhostar}, the erroneous $\rho_*$ value really comes from an
erroneous $a/R_*$ value. Since the derived orbital period will be reliable
irrespective of the orbital eccentricity, the only cause of $\rho_*$ deviating 
from the true value is because $(a/R_*)^3$ deviates from the true value. Using a 
set of approximate expressions for the transit duration, which are accurate to 
99.9\% for planets of $|e\sin\omega|<0.5$ and $|e\cos\omega|<0.85$,
\citet{investigations:2010} showed that if one assumes a circular orbit then the 
derived value $(a/R_*)$ differs from the true value via:

\begin{align}
[(a/R_*)^{\mathrm{circ}}]^2 = \frac{(1+p)^2 - [b^{\mathrm{circ}}]^2}{\sin^2 (T_{1,4} \pi/P)} + [b^{\mathrm{circ}}]^2
\end{align}

where $b^{\mathrm{circ}}$ is given by:

\begin{align}
&[b^{\mathrm{circ}}]^2 = 1 + p^2 + 2p \nonumber \\
&\Bigg[ \Bigg( \sin^2[\frac{\varrho_c^2}{\sqrt{1-e^2}} \arcsin(\frac{\sqrt{(1-p)^2 - b^2}}{(a/R_*) \varrho_c \sin i})] \nonumber \\
&+ \sin^2[\frac{\varrho_c^2}{\sqrt{1-e^2}} \arcsin(\frac{\sqrt{(1+p)^2 - b^2}}{(a/R_*) \varrho_c \sin i})] \Bigg) \nonumber \\
&\Bigg( \sin^2[\frac{\varrho_c^2}{\sqrt{1-e^2}} \arcsin(\frac{\sqrt{(1-p)^2 - b^2}}{(a/R_*) \varrho_c \sin i})] \nonumber \\
&- \sin^2[\frac{\varrho_c^2}{\sqrt{1-e^2}} \arcsin(\frac{\sqrt{(1+p)^2 - b^2}}{a_R \varrho_c \sin i})] \Bigg)^{-1} \Bigg]
\end{align}

where $\varrho_c=(1-e^2)/(1+e\sin\omega)$ and $p$ is the ratio-of-radii. The 
relative difference between our approximate expression for the stellar density 
(i.e. the $\Psi$-equation) and the more accurate value is therefore given by the 
LHS of the following:

\begin{align}
\frac{[(a/R_*)^{\mathrm{circ}}]^3 - (a/R_*)^3 \Psi}{[(a/R_*)^{\mathrm{circ}}]^3} < t
\end{align}

Where $t$ is the tolerance level for the desired level of accuracy. For example,
a typical choice might be $t=10^{-3}$ indicating 99.9\% accuracy in the
$\Psi$-equation. For brevity, we do not write out the full form of the above
expression. It is trivial to show that it is maximized for $\omega=\pi/2$ and
$b=1$ and therefore if we satisfy a given tolerance level under these 
conditions then we can be sure the equation is always valid. Eliminating these
two terms accordingly, one may then take the limit of the resulting equation
for when $p\rightarrow0$, corresponding to the small-planet approximation which
is essentially valid for $p\lesssim0.1$, encompassing almost all transiting
planets. This leads to the far simpler expression:

\begin{align}
& \Bigg| (a/R_*)^3 \Bigg(\frac{(1+e)^3}{(1-e^2)^{3/2}}\Bigg) \nonumber \\
\qquad& \Bigg(\frac{(a/R_*)^2 (1-e)^2 (1+e) - e(4-3e+e^2)}{(1-e)^3}\Bigg)^{-3/2} \Bigg| > 1 - t
\end{align}

\begin{figure}
\begin{center}
\includegraphics[width=8.4 cm]{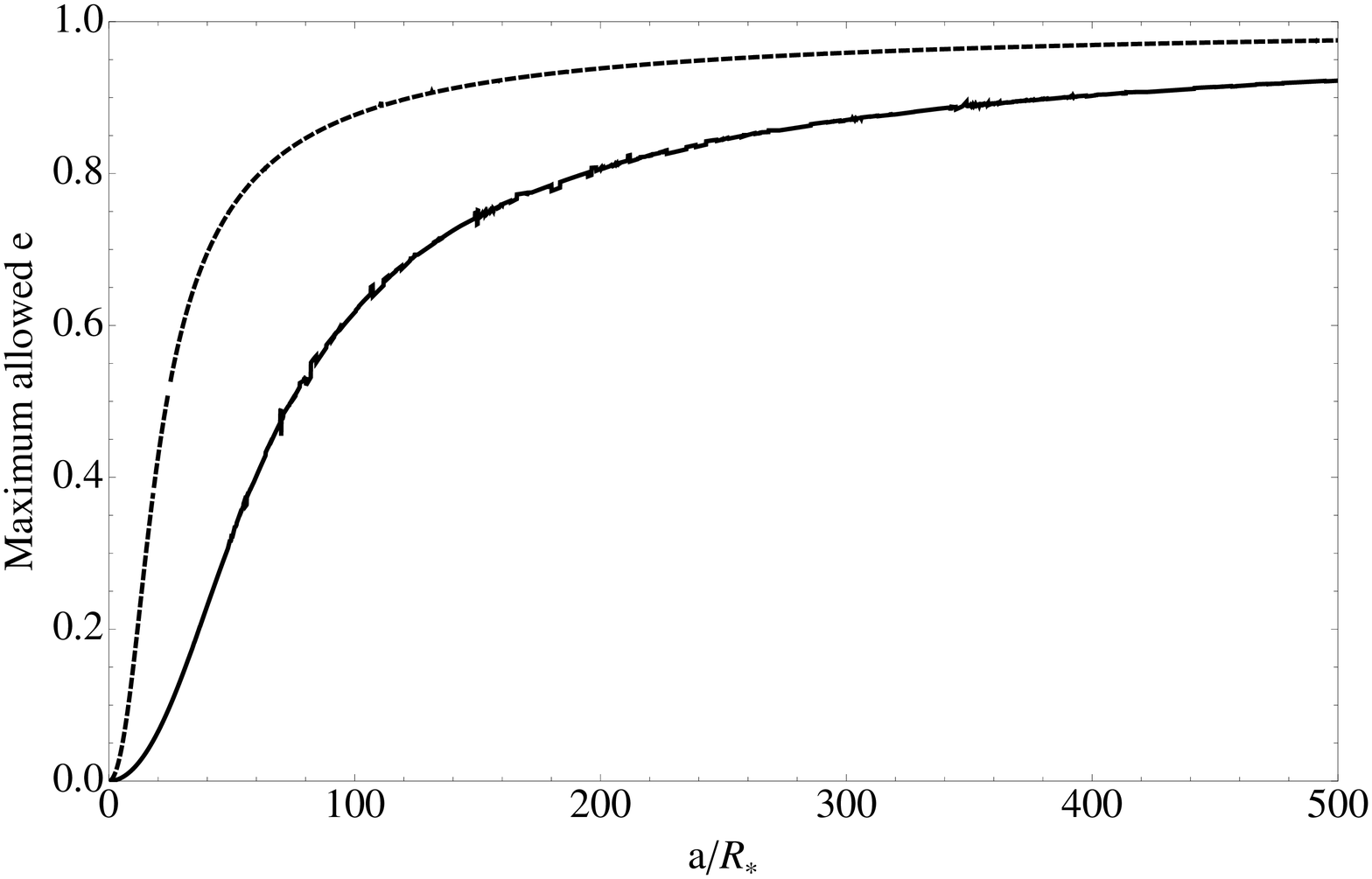}
\caption{\emph{Maximum allowed orbital eccentricity versus $(a/R_*)$ in
order for the $\Psi$-equation to remain a valid approximation. The two lines
correspond to two accuracy tolerance levels. The dashed line is for 99\%
accuracy and the solid is for 99.9\%. Small bumps are due to numerical errors.}} 
\label{fig:maxecc}
\end{center}
\end{figure}

This expression indicates that our accuracy becomes worst for low $(a/R_*)$ and 
high $e$ values. Numerically solving for the maximum allowed $e$ as a function
of $(a/R_*)$ may be accomplished for a given $t$ level to illustrate some
typical constraints. In Figure~\ref{fig:maxecc}, we show the case for 
$t=10^{-2}$ (dashed) and $t=10^{-3}$ (dotted).

Since even the highest precision measurement uncertainty on $\rho_*$ is around 
1\% (e.g. \citealt{kippingbakos:2011b}), $t=10^{-2}$ is sufficient for our 
purposes. For some typical values of $(a/R_*)$ of $(a/R_*) = $10, 100, 300 \& 
1000 (corresponding to a hot-, warm-, temperate- and cold-planet respectively, 
assuming a Solar-star) we find $e_{\mathrm{max}} = $0.15, 0.88, 0.96 \& 0.99 
respectively. If these limits are exceeded, the techniques presented later in 
this paper will still detect orbital eccentricity but the actual determination 
of the $e$ will obviously be subject to a systematic error.

This limitation may seem to be a significant disadvantage but in reality one
does not expect to find multiple planet systems with very high eccentricities.
The most feasible scenario where the expressions would become invalid
is for a multiple planet system featuring one very close-in planet. However,
in such a case, one would not expect the planet to be highly eccentric anyway
due to rapid tidal circularization at such distances. Nevertheless, if the 
planet is both close-in and highly eccentric, one can simply discount the planet
during the MAP analysis.


\section{MAP: Numerical Constraints (\MakeLowercase{n}-MAP)}
\label{sec:n-MAP}

\subsection{Description of the Algorithm}
\label{sub:algorithm}

Whilst the approximate inequalities derived in \S\ref{sec:a-MAP} are useful, 
they do not take advantage of the extra information offered by $>2$ planets
and are invalid at large eccentricities since we are ignoring terms of 
$\mathcal{O}[e_k^2]$ or larger. To address this, one may consider a more 
brute-force approach through numerical methods.

\S\ref{sec:a-MAP} has shown that converting 
$\Theta_{ij} \rightarrow \{e_k,\omega_k\}$ is degenerate and in general we
can only obtain some lower bounds on $e_k$. However, it is trivial to perform
the reverse operation and convert $\{e_k,\omega_k\}\rightarrow\Theta_{ij}$.
Therefore, although computationally demanding, one could create a 
multi-dimensional grid of all possible $e_k$ and $\omega_k$ values, for all 
planets, and compute the $\Theta_{ij}$ values at each grid point. The
$\Theta_{ij}$ value at each grid point could then be compared to the observed 
values of $\Theta_{ij}$ to evaluate the likelihood of the $\{e_k,\omega_k\}$ 
vector at the specified grid location. This likelihood can then be used to 
create maps of the permitted/excluded parameter regions for $\{e_k,\omega_k\}$. 
This numerical approach is an extension to the analytic approximations used 
earlier. Both techniques can be regarded as what we call ``Multibody 
Asterodensity Profiling'' (MAP). We distinguish between the two approaches as 
analytic-MAP (a-MAP) and numerical-MAP (n-MAP).

Although we have just described a grid search in the previous paragraph, a much
more computationally efficient and powerful numerical technique is to
adopt a Markov Chain Monte Carlo (MCMC) algorithm. Our overall approach to 
numerical-MAP utilizes two-stages of a Metropolis-Hastings MCMC fitting routine. 
The first is for the light curve fits and the second is performed once the light 
curve MCMCs are complete. Experimentation with combining the two stages into one 
larger MCMC yielded inordinately large computation times, whereas the two-stage 
technique provides results in just a few hours on a typical workstation. Our 
method can be loosely described via the following algorithm:\\

\textbf{Stage 1}
\begin{itemize}
\item[\textbf{1.1}] For an $n$-body system, fit transit light curves for each 
planet independently, assuming a circular orbit, using a Markov Chain Monte 
Carlo (MCMC) algorithm with the Metropolis-Hastings rule.
\item[\textbf{1.2}] Compute $n!/[2!(n-2)!]$ marginalized posterior distributions 
for $\Theta_{ij}$ for $i=1,2,...n$, $j=1,2,...n$ and $i\neq j$. Each posterior is
ensured to have $A$ down-sampled points from a well-mixed and converged chain.
\item[\textbf{1.3}] Normalize the posteriors such that they become equal to
the probability density function (PDF) of each $\Theta_{ij}$. We dub these
PDFs as $\mathrm{PDF}(\Theta_{ij})$.
\end{itemize}
\textbf{Stage 2}
\begin{itemize}
\item[\textbf{2.1}] Define a starting point for a new MCMC chain with a fitting
parameter set $\vec{\Lambda_b} = \{e_{k,b},\omega_{k,b}\}$ for 
$k=1,2,...n$, where $b$ is understood to represent the $b^{\mathrm{th}}$
accepted MCMC trial.
\item[\textbf{2.2}] Using Equation~\ref{eqn:theta}, evaluate $\Theta_{ij,b}$ 
for $i=1,2,...n$, $j=1,2,...n$ and $i\neq j$.
\item[\textbf{2.3a}] If the trial value of 
$\Theta_{ij,b}>\mathrm{Mode}[\mathrm{PDF}(\Theta_{ij})]$, then count the number
of realizations in the $\mathrm{PDF}(\Theta_{ij})$ which fall in the range of
$\mathrm{Mode}[\mathrm{PDF}(\Theta_{ij})]<\Theta_{ij}<\Theta_{ij,b}$ and define
this integer as $m_{ij,b}$.
\item[\textbf{2.3b}] If the trial value of 
$\Theta_{ij,b}<\mathrm{Mode}[\mathrm{PDF}(\Theta_{ij})]$, then count the number
of realizations in the $\mathrm{PDF}(\Theta_{ij})$ which fall in the range of
$\Theta_{ij,b}<\Theta_{ij}<\mathrm{Mode}[\mathrm{PDF}(\Theta_{ij})]$ and define
this integer as $m_{ij,b}$.
\item[\textbf{2.4}] Define the $\chi^2$ of the $b^{\mathrm{th}}$ MCMC trial
as $\sum_{i\neq j}^n \chi_{ij,b}^2$ where 
$\chi_{ij,b}^2 = (\sqrt{2} \mathrm{Erf}^{-1}[2m_{ij,b}/A])^2$.
\item[\textbf{2.5}] Accept/Reject trial point following the Metropolis-Hastings
rule and loop the MCMC in the usual manner until $B$ trials have been accepted.
\end{itemize}

By the end of the algorithm, we have obtained a joint-posterior for the 
$\{e_k,\omega_k\}$ vector revealing those regions of parameter space which are
excluded and those which are more probable. The merit function (inverse of
the likelihood) is thus given by:

\begin{align}
\chi_{\mathrm{MAP}}^2 &= \sum_{i\neq j}^n \Bigg( \frac{ \Theta_{ij,\mathrm{obs}} - \Theta_{ij,\mathrm{model}} }{ \sigma(\Theta_{ij}) } \Bigg)^2
\label{eqn:chi2}
\end{align}

where it is understood that $\sigma$ is determined by numerically integrating
the probability density function of $\Theta_{ij,\mathrm{obs}}$ (steps 2.3\&2.4), 
rather than simply counting the number of 1$\sigma$ error bars between the model 
and observed value of $\Theta_{ij}$. The advantage of doing this is that we
are able to fully account for non-Gaussian $\Theta_{ij,\mathrm{obs}}$ 
posteriors, which are common as will become evident in \S\ref{sub:KOI-S01}, 
\ref{sub:KOI-S02} \& \ref{sub:KOI-S0C}. The disadvantage is significantly
increased computation time since each MCMC trial of n-MAP requires 
$\sim n!/[2!(n-2)!] A$ evaluations. 

We also point out that one may add additional information into the technique
at this stage. For example, if radial velocity or TTV information exists and
can be used to place further constraints on $e_k$, then additional components
to the total merit function can suitably appended. However, for the remainder
of this work we focus on MAP alone to demonstrate the use of this technique
as a unique type of observable.

To compute the inverse error function, we use the approximation of 
\citet{winitzk:2006}, which is accurate to $4\times10^{-3}$ 
over the interval $0<x<1$:

\begin{align}
\mathrm{Erf}^{-1}(x) &\simeq \Bigg[ -\frac{2}{\pi \beta} - \frac{\log(1-x^2)}{2} \nonumber \\ 
\qquad& + \sqrt{\Big(\frac{2}{\pi \beta} + \frac{\log(1-x^2)}{2}\Big)^2 - \frac{1}{\beta}\log(1-x^2)} \Bigg]^{1/2}
\end{align}

where $\beta = [8(\pi-3)]/[3\pi(4-\pi)]$. 
In the next subsection we discuss how we enforce a uniform prior in 
$(\rho_{*,k}^{\mathrm{circ}})^{2/3}$, which enables a uniform prior in 
the $\Theta_{ij}$ posteriors.

\subsection{Light Curve Fitting Parameters}
\label{sub:fittingparams}

The choice of fitting parameters for the transit light curve has a
broad diversity within the exoplanet literature and yet a
significant impact on the derived results and efficacy of a light curve
fitting algorithm. In this subsection, we will describe the light
curve fitting parameter set which yields the most reliable results for the
specific purposes of MAP. In order to accomplish this, we will briefly overview
the basic properties of the light curve.

\subsubsection{Understanding the trapezoid light curve}

The transit light curve is well-approximated by a trapezoid in the limit of
negligible limb darkening. A trapezoid is described by four parameters only: 
$T_{14}$, the duration from the $1^{\mathrm{st}}$-to-$4^{\mathrm{th}}$ contact; 
$T_{23}$, the duration from the $2^{\mathrm{nd}}$-to-$3^{\mathrm{rd}}$ contact; 
$\delta$, the depth of the trapezoid and $t_{\mathrm{mid}}$, the mid-point of 
the trapezoid in time. The parameter set may be written as 
$\vec{\Gamma} = \{\delta,T_{14},T_{23},t_{\mathrm{mid}}\}$. It is easy to see 
that one could alternatively use the ingress/egress duration ($T_{12}$) and the 
full-width-half-maximum duration ($W$) instead of $T_{14}$ and $T_{23}$.

\subsubsection{Understanding the circular orbit light curve}

Consider a transiting planet on a circular orbit but with limb darkening on
the star. It is easy to show that four parameters only can still be used to
completely describe the light curve, even though the light curve is no longer
morphologically trapezoidal \citep{thesis:2011}. These terms are: $(a/R_*)$,
the semi-major axis of the planetary orbit around the star in units of the
stellar radius; $b$, the sky-projected distance between the planet and the star
at the instant of inferior conjunction in units of the stellar radius (often
called the impact parameter); $p^2$, the square of the ratio of the planet's
radius to the stellar radius and $\tau$, the instant when the sky-projected
planet-star separation is minimized in proximity to the instant of inferior
conjunction. So we have $\vec{\Gamma} = \{p^2,(a/R_*),b,\tau\}$. 

One can easily appreciate that $p^2$ replaces $\delta$ (but are equivalent for a 
non-limb darkened star) and $\tau$ replaces $t_{\mathrm{mid}}$ (but are 
equivalent for circular orbits). Further, $\{(a/R_*),b\}$ replace
$\{T_{14},T_{23}\}$. However,as shown by \citet{seager:2003}, these terms
are interchangeable via:

\begin{align}
\lim_{e \rightarrow 0} b &= \Bigg[ \frac{(1-p)^2-\frac{\sin^2(T_{23} \pi/P)}{\sin^2(T_{14} \pi/P)} (1+p)^2}{1-\frac{\sin^2(T_{23} \pi/P)}{\sin^2(T_{14}\pi/P)}} \Bigg]^{1/2} \\
\lim_{e \rightarrow 0} (a/R_*) &= \Bigg[ \frac{(1+p)^2-[b^2(1-\sin^2(T_{14}\pi/P)}{\sin^2(T_{23}\pi/P)}\Bigg]^{1/2}
\label{eqn:seagerparams}
\end{align}

Therefore, one has the choice as to whether one uses $\{(a/R_*),b\}$
or $\{T_{14},T_{23}\}$. Indeed, one can also legitimately use many
other combinations which are interchangeable, such as $\{W,T_{12}\}$
\citep{carter:2008}, $\{\tilde{T},T_{12}\}$ \citep{investigations:2010},
$\{(\zeta/R_*),b^2\}$ \citep{bakos:2007}, $\{(\Upsilon/R_*),b^2\}$ 
\citep{investigations:2010}, etc. This already 
raises the question as to what parameter set should be used. The two terms
are problematic in that they typically exhibit mutual correlation and so
care must be taken in their selection.

\subsubsection{Understanding the eccentric orbit light curve}

For an eccentric orbit, the morphology of the light curve is essentially
unchanged. The signal of asymmetry is negligible and will rarely affect
measurements for even extreme cases \citep{kipping:2008,winn:2010}. As a result,
the eccentric terms $e$ and $\omega$ (eccentricity and position of pericentre)
are hidden from view and cannot be determined by simply fitting a light curve 
(obviously, for multi-planet systems an alternative, more subtle strategy exists 
in the form of MAP). Consequently, the same parameter set applies for eccentric 
orbits as for circular orbits i.e. one can use, for example, 
$\vec{\Gamma} = \{p^2,(a/R_*),b,\tau\}$. The only difference is that the we have
to declare values for $e$ and $\omega$ during the fits. These eccentric terms
do affect the relationship between $\{(a/R_*),b\}$ and duration related terms
and a modified form of Equation~\ref{eqn:seagerparams} should be used, as
presented in \citet{investigations:2010}. Indeed, it is these differences
which fundamentally allow MAP to work.

\subsubsection{Choosing a parameter set}

In this work, the term which we are interested in is the derived value of
$\Theta_{ij}$, which has been established to contain information about the 
orbital eccentricity. This term is simply the ratio of 
$(\rho_*^{\mathrm{circ}})^{2/3}$ values. As is discussed in 
\S\ref{sub:indirectpriors}, uniform priors in the eccentricity terms for n-MAP 
can be implemented by ensuring a uniform prior in 
$(\rho_*^{\mathrm{circ}})^{2/3}$. By fitting for, say 
$\{p^2,(a/R_*),b,\tau_i\}$, we necessarily assume uniform priors on those terms. 
However, since $(\rho_*^{\mathrm{circ}})^{2/3}$ is an intricate function of 
these terms, it will not have a uniform prior.

A simple but effective solution to this problem is to fit for 
$(\rho_*^{\mathrm{circ}})^{2/3}$ directly. This term may be easily converted to
$(a/R_*)$ via:

\begin{align}
(a/R_*) & = \Bigg( \frac{G P^2 \rho_{*}^{\mathrm{circ}} }{ 3\pi } \Bigg)^{1/3}
\end{align}

This leaves one of the two problematic terms assigned, but still leaves us
with options for the other. For example, should we use 
$\{p^2,(\rho_*^{\mathrm{circ}})^{2/3},b,\tau_i\}$, 
or $\{p^2,(\rho_*^{\mathrm{circ}})^{2/3},T_{12},\tau_i\}$,
or $\{p^2,(\rho_*^{\mathrm{circ}})^{2/3},\cos(i),\tau_i\}$
or many other possible permutations?

\subsubsection{The other term}

Choosing this term is non-trivial. If for example, we chose $b$, we would be
faced with the issue that $b<0$ is unphysical and thus a boundary condition
exists at $b=0$. In a Markovian sense, jumps to negative $b$ values are rejected
thus resulting in a disequilibrium between the number of positive and negative
jumps. This in turn means that the $b$ posterior will be biased and overestimate 
the true value. Since inter-parameter correlations exist between the
light curve fitting parameters, a bias in $b$ induces a bias in 
$(\rho_*^{\mathrm{circ}})^{2/3}$. Terms being correlated in itself is not
a major problem, it slows down our algorithm but an accurate result can still
be obtained with a sufficient number of trials. However, if one of the
correlated terms is biased then the fact the terms are correlated to one another
becomes a problem, since now all terms will become biased.

The boundary condition in $b$ in therefore a serious issue. A similar situation 
is well-known to exist for $e$ with radial velocity fits \citep{lucy:1971}.
One could propose that using a duration-based term such as $T_{14}$, $\tilde{T}$
or $T_{12}$ would avert such a problem. However, for certain duration jumps, the 
derived impact parameter still falls out as being unphysical (in this case
imaginary). These unphysical trials can be discarded but that again
introduces a bias. Therefore, any other duration related parameter would also be 
a boundary-condition-limited parameter.

To allow $b$ to make Markovian steps, we let $b$ go to negative values. Since
the light curve and duration-related terms are always generated using $b^2$,
then the physicality of $b<0$ solution is irrelevant mathematically speaking.
We found that this yielded solutions consistent with test cases and thus
seems to solve the problem.

In addition to $\{p^2,(\rho_*^{\mathrm{circ}})^{2/3},b,\tau_i\}$, we also
allow each transit epoch to have a unique out-of-transit normalization factor,
$OOT_m$ where $m$ denotes the epoch number. The zeroth epoch is defined to be
that which has the lowest mutual correlation to the orbital period. Finally,
the orbital period is a free parameter too.

\subsection{Direct \MakeLowercase{n}-MAP Priors}
\label{sub:directpriors}

In most cases (system with less than five transiting planets), there are more
free parameters in the model than observables and so the problem is 
under-constrained with no unique solution. Despite this, contours of the
error surface may still be computed through Monte Carlo techniques. In our
case, the Monte Carlo technique of choice is Markov Chain Monte Carlo (MCMC)
with the Metropolis-Hastings algorithm. We note that an analogous situation
arises for interpreting the atmospheres of exoplanets where atmosphere
models tend to have more free parameters than the number of observations.
In this example, a similar solution as n-MAP has been adopted in works
such as \citet{madhu:2009} and \citet{madhu:2011}. A direct comparison of
n-MAP to the methods of \citet{madhu:2011} is discussed in 
\S\ref{sub:madhucomparison}.

We stress here that because the problem is under-constrained (except for 
$n\geq5$), the results will be strongly affected by the choice of priors. This 
is in contrast to a highly constrained problem where the data drive the result 
to the same solution with only a minor dependency upon the choice of priors.
Thus, the choice of priors can be understood to have a significant
impact on the derived results (see \citealt{ford:2005} for a detailed
discussion on the effect of priors when fitting exoplanet data). The results
should consequently be always quoted in unison with the adopted prior used to 
infer them.

These ideas are more formally expressed through Bayes' theorem. Let us denote
the eccentricity parameters which we fit for in n-MAP as 
$\vec{\Lambda} = \{e_k,\omega_k,...\}$ for $k=1,n$. We further use
$\mathcal{M}$ to represent the model and $\mathcal{D}$ to represent the data:

\begin{align}
\mathrm{P}(\vec{\Lambda}|\mathcal{D},\mathcal{M}) &= \frac{ \mathrm{P}(\vec{\Lambda}|\mathcal{M}) \mathrm{P}(\mathcal{D}|\vec{\Lambda},\mathcal{M}) }{ \int \mathrm{P}(\vec{\Lambda}|\mathcal{M}) \mathrm{P}(\mathcal{D}|\vec{\Lambda},\mathcal{M})\,\mathrm{d}\vec{\Lambda} }
\label{eqn:bayes1}
\end{align}

In our case, the ``data'' is the observed ratios of 
$(\rho_{*,i}^{\mathrm{circ}}/\rho_{*,j}^{\mathrm{circ}})^{2/3}$, denoted by the 
term $\Theta_{ij}$:

\begin{align}
\mathrm{P}(\vec{\Lambda}|\vec{\Theta},\mathcal{M}) &= \frac{ \mathrm{P}(\vec{\Lambda}|\mathcal{M}) \mathrm{P}(\vec{\Theta}|\vec{\Lambda},\mathcal{M}) }{ \int \mathrm{P}(\vec{\Lambda}|\mathcal{M}) \mathrm{P}(\vec{\Theta}|\vec{\Lambda},\mathcal{M})\,\mathrm{d}\vec{\Lambda} } \nonumber \\
\mathrm{P}(\vec{\Lambda}|\vec{\Theta},\mathcal{M}) &= \frac{ \mathrm{P}(\vec{\Lambda}|\mathcal{M}) \mathrm{P}(\vec{\Theta}|\vec{\Lambda},\mathcal{M}) }{ \mathrm{P}( \vec{\Theta} )}
\label{eqn:bayes2}
\end{align}

Given that the problem is under-constrained, clearly the 
choice of this prior will have a significant impact on the derived joint
probability distribution from n-MAP. There are two plausible paths to adopt:

\begin{itemize}
\item[{\tiny$\blacksquare$}] Assume complete ignorance for the a-priori 
knowledge of $\vec{\Lambda}$
\item[{\tiny$\blacksquare$}] Adopt a prior based upon dynamics and/or known 
prior distribution of eccentricity
\end{itemize}

For the former, complete ignorance can be easily implemented by adopting a
uniform prior in $\vec{\Lambda}$. This would take the form of a uniform prior
between $0 \leq e_k < 1$ and $0 \leq \omega_k < 2\pi $. One advantage of this
choice is that any results derived from n-MAP can be understood to be directly
due to the MAP technique rather than any prior biases. One disadvantage is
that we know that a system of multiple planets is unlikely to survive with high 
eccentricities and we are essentially ignoring this fact. However, this could
also be considered a potential advantage in that a system where n-MAP strongly 
prefers a dynamically unstable solution may indicate that the system is in 
fact a false positive.

For the second option, a typical procedure is outlined in the recent work of 
\citet{steffen:2010} for five candidate multiple transiting planet systems 
detected by \emph{Kepler}. Here, the authors adopted a prior distribution
in $e_k$ based upon the same distribution discussed earlier in 
\S\ref{sub:aMAPreliability} i.e. a mixture of an exponential and a Rayleigh 
distribution following \citet{juric:2008} and \citet{zakamska:2011} and provided
earlier in Equation~\ref{eqn:eprior}.

As already touched on, in this work we prefer to present n-MAP results
with uniform priors in $\vec{\Lambda}$ (i.e. assume total ignorance) for the
sake of demonstrating this new technique, but future works could make use of
more sophisticated priors like those of \citet{steffen:2010}. By making this 
choice, the derived constraints in this work can be understood to be completely 
due to the n-MAP method alone, and not due to the impact of priors. 

With this choice, it is arguably better to think about the n-MAP
results in terms of ``allowability-space'' rather than probability space. An 
area of high-density from n-MAP signifies a region where lots of combinations of 
parameters can reproduce $\rho_{*,k}^{\mathrm{circ}}$'s which agree well with 
the data. An area of low-density signifies a region where very few combinations 
of parameters can reproduce $\rho_{*,k}^{\mathrm{circ}}$'s which agree with the 
data. An area of null-density signifies that absolutely no combination of 
parameters can reproduce the observed $\rho_{*,k}^{\mathrm{circ}}$'s.

\subsection{Indirect \MakeLowercase{n}-MAP Priors}
\label{sub:indirectpriors}

Aside from the priors in the MCMC chain of the n-MAP phase, priors also
affect n-MAP indirectly via the light curve fits. If we, for
example, fit for $\rho_{*,k}^{\mathrm{circ}}$ in the light curve fits, the prior
on $\Theta_{ij}$ will be non-uniform. Even if $\Theta_{ij}$ has a uniform
prior, it is not immediately obvious that this will translate to uniform priors 
in $e_k$ and $\omega_k$.

Consider first that one executes n-MAP with the $\Theta_{ij}$ terms behaving as
uniformly distributed parameters. $\Theta_{ij}$ is used to compute the $\chi^2$ 
of the n-MAP realizations in step 2.3 (as described in \S\ref{sub:algorithm}). 
This calculation requires that we know the mode of $\Theta_{ij}$, which is a 
meaningless concept for a uniform distribution. Nevertheless, one can easily 
appreciate that the $\chi^2$ must be the same for all MCMC realizations of 
$\{e_k,\omega_k\}$, since $\Theta_{ij}$ is uniformly distributed. In
therefore follows that all MCMC realizations will be accepted under the
Metropolis-Hastings rule. If all trials are accepted, then this is equivalent
to the case KOI-S0P described in \S\ref{sub:KOI-S0P}, which simply
reproduces the behaviour of the direct priors i.e. uniform in 
$\{e_k,\omega_k\}$. Consequently, uniform priors in $\Theta_{ij}$ are something
to be desired since it does not cause any bias in the resulting n-MAP procedure.

With this point established, the next question is how can we ensure uniform
priors are produced for $\Theta_{ij}$? Since $\Theta_{ij}$ is the ratio of
$(\rho_{*,i}^{\mathrm{circ}})^{2/3}/(\rho_{*,j}^{\mathrm{circ}})^{2/3}$, one
first step would be to adopt uniform priors in 
$(\rho_{*,k}^{\mathrm{circ}})^{2/3}$. However, the ratio of two uniform
priors does not produce a uniform prior itself, rather we have:

\begin{equation}
\label{eqn:uniformratio}
\mathrm{P}(\Theta_{ij}) = 
\left\{\begin{array}{ll}
\frac{1}{2}  &  0 < \Theta_{ij} \le 1 \\
\frac{1}{2 \Theta_{ij}^2} &  \Theta_{ij} > 1 \\
0 & \mathrm{otherwise} \\
\end{array}\right.
\end{equation}

Ergo, by fitting the transit light curves for 
$(\rho_{*,k}^{\mathrm{circ}})^{2/3}$, we adopt a uniform prior in this term but
the derived $\Theta_{ij}$ terms will be non-uniform for $\Theta_{ij}>1$. Our
solution for tackling this is to adapt the n-MAP algorithm. The solution lies
in the fact that $\Theta_{ij}$ is uniform for values $<1$. If we generate a
$\{e_k,\omega_k\}$ realization which causes $\Theta_{ij}>1$, we may simply
use $\Theta_{ji}$ instead, which must lie in the range $0<\Theta_{ji}<1$
and therefore must be uniformly distributed according to 
Equation~\ref{eqn:uniformratio}. By implementing this condition, we ensure
that both the direct and indirect priors in n-MAP are uniform in 
$\{e_k,\omega_k\}$.

\subsection{Difference Between \MakeLowercase{n}-MAP and the \citet{madhu:2011} 
Technique}
\label{sub:madhucomparison}

There are several methodological similarities between the n-MAP 
technique to constrain orbital eccentricity and the numerical methods proposed 
by \citet{madhu:2009}, and 
subsequent papers, to determine exoplanet atmospheric composition. In these
papers, the authors produced parameter space maps showing the points which
agree to the data to within $\chi^2<1$, $\chi^2<2$ and $\chi^2<3$ to denote
different error surfaces. As of \citet{madhu:2011}, MCMC methods were used for
the parameter space exploration rather than grid methods which were used 
in \citet{madhu:2009}. Despite switching to MCMC, the presentation of results
remained largely unchanged with $\chi^2<1$, etc surfaces still being plotted. 
Such a presentation does not take advantage of the fact the MCMC technique 
inherently computes the probability density over the parameter space, rather 
than merely outputting the likelihood of individual realizations. To accomplish 
this, one simply computes the parameter space regions in which the MCMC spends 
the majority of its time. These regions represent the high probability density
areas. Therefore, we can see that there are two possible paths by which to 
proceed.

The analogy between n-MAP and the method of \citet{madhu:2011} breaks down here.
For us, the first way of presenting the results would be a far less useful
diagnostic of the parameter space. This is because almost any $e_i$-$e_j$
combination can be found to give a $\chi^2<1$ for the correct tuning of
$\omega_i$ and $\omega_j$. However, the fine tuning of these parameters
must be so precise, that very few MCMC realizations find such values.
Nevertheless, we typically compute $\mathcal{O}[10^6]$ points in the MCMC chain
and so these improbable locations will eventually be visited by the chain.
Consequently, if we plotted the minimum $\chi^2$ in a rasterized grid of
$e_i$-$e_j$, analogous to the presentation in \citet{madhu:2011}, we would 
essentially find a region where any solution is permitted. In contrast, plotting 
the probability density regions in $e_i$-$e_j$ space automatically accounts for 
the fact that despite these locations yielding a low $\chi^2$, they require very 
precise tunings of $\omega_i$ and $\omega_j$. This is the key difference in the 
presentation of our results.

A way to visualize this more easily to consider the result we would obtain
for a two planets on a circular orbit. An numerical example will illuminate
this issue more fully. Let us say 
$\rho_{*,1}^{\mathrm{circ}} = 1.00\pm0.0071$\,g\,cm$^{-3}$ and
$\rho_{*,2}^{\mathrm{circ}} = 1.00\pm0.0071$\\,g\,cm$^{-3}$, which is consistent 
with that which would be obtained for two planets on circular orbits. These
derived stellar densities would yield $\Theta_{12}=1.000\pm0.0067$. Now consider 
that during the MCMC parameter exploration, one realization is attempted where 
$e_1 = 0.9$ and $e_2 = 0.4$. This would seem to be a position that one would 
expect to be highly excluded by the n-MAP technique. We have:

\begin{align}
\Theta_{12} &= \Bigg(\frac{1+e_1 \sin \omega_1}{1+e_2 \sin \omega_2}\Bigg)^2 \Bigg(\frac{1-e_2^2}{1-e_1^2}\Bigg)
\end{align}

Solving for $\Theta_{12} = 1$ for $\omega_2$, we have:

\begin{align}
\omega_2 &= \arcsin\Bigg[\frac{e_2 (1-e_1^2)-\sqrt{e_2^2 (1-e_1^2)(1-e_2^2) (1+e_1\sin\omega_1)^2}}{-e_2^2 (1-e_1^2)}\Bigg]
\end{align}

\begin{figure}
\begin{center}
\includegraphics[width=8.4 cm]{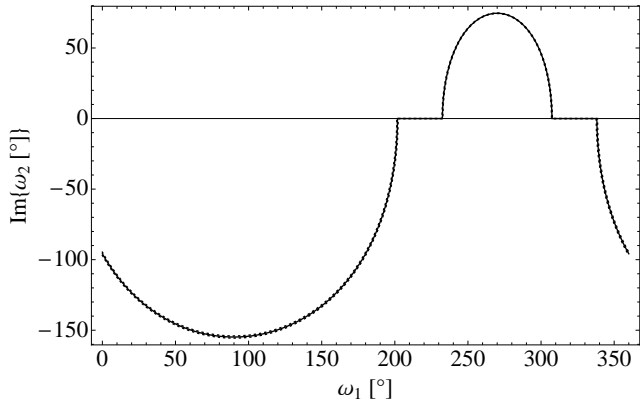}
\caption{\emph{
For $e_1=0.9$ and $e_2 = 0.4$, yet $\Theta_{12} = 1.00$, the above
shows the range of the imaginary component of $\omega_2$ values which allow
a solution of $\chi^2=0$ in solid and $\chi^2=9$ in dotted, as a function of 
$\omega_1$. Solutions requiring non-zero imaginary $\omega_2$ can never be 
reached by the n-MAP algorithm. However, a narrow range of $\omega_1$ allows 
both real $\omega_1$ and $\omega_2$ solutions. For this reason, although 
improbable, n-MAP can find a low $\chi^2$ solution at virtually all $e_i-e_j$ 
coordinates. However, these states are visited with a low frequency due to the 
above constraints.}} 
\label{fig:imagin}
\end{center}
\end{figure}

Plotting the imaginary component of this equation as a function of $\omega_1$
for the fixed values of $e_1 = 0.9$ and $e_2 = 0.4$ (Figure~\ref{fig:imagin}), 
a narrow range of $\omega_1$ values can still yield a real result. Specifically, 
in this example, 17.2\% of the allowed $\omega_1$ range can yield a solution. 
One solution occurs at $\omega_1=217^{\circ}$. We can therefore imagine the
MCMC algorithm landing upon this value too, thus permitting a physical solution. 
Even if it happens to land within this 17.2\% sliver of $\omega_1$, we also 
require a very narrow range of $\omega_2$. Here, we require 
$\omega_1 = (354.8\pm0.5)^{\circ}$ in order to obtain $\chi^2<1$ in n-MAP. This 
is a range of just 0.3\% of all possible $\omega_2$ values.

Putting this all together, for a genuinely circular orbit system, a random 
realization of $e_1=0.9$ and $e_2=0.4$ can still have some n-MAP trials 
of $\chi^2<1$. However, the fraction of times which n-MAP will succeed in doing
this will be less than 0.05\% (for uniform priors in $\omega_k$). Nevertheless, 
for long chain lengths we should expect n-MAP to find $\chi^2<1$ at virtually 
every single point in $e_i$-$e_j$ parameter space. Therefore, presenting plots 
of points where realizations yielded various $\chi^2$ is of limited value. What 
is much more useful, is to plot the density of realizations across the 
$e_i$-$e_j$ parameter space resulting from n-MAP. This inherently accounts for 
the fact certain $e_i$-$e_j$ combinations require fine tunings of the other 
terms in order to produce an acceptable trial. This is the chief difference 
between n-MAP and the method of \citet{madhu:2011}.

\subsection{MCMC Diagnostics}
\label{sub:mcmcdiagnostics}

We will later show applications of our algorithm to synthetic data sets. In
these cases, it is important to ensure that the MCMC fits for both the
transit light curves and n-MAP achieve i) adequate mixing 
ii) adequate convergence. 

\subsubsection{Burn-in}

Before either of these diagnostics can be computed,
it is important to remove the pre-burn trials of the MCMC. These trials
are highly dependent upon the initial starting point of the chain and thus it
is important to burn-out the initial part of the chain. We will use the same
strategy as \citet{tegmark:2004} for this. Specifically, we compute the median
$\chi^2$ of all accepted MCMC trials and then burn-out the initial trials up
to the point when the $\chi^2$ drops below the median value. Burn-out is
typically very rapid and occurs within a few dozen trials.

\subsubsection{Mixing}

To determine whether our chains are sufficiently mixed, we compute the effective
length of the chain (see \citealt{tegmark:2004}). Each free parameter has its
own unique effective length and so we always conservatively adopt the lowest
effective length in reporting the final diagnostics. Broadly speaking, we wish
to reach a point where the lowest effective length $\gg 1$, such that 
meaningful statistics can be inferred. In this work, we set the goal that the 
lowest effective length $\gtrsim 1000$. This is sometimes achieved by combining 
multiple chains rather than simply extending the length of a single chain. The 
advantage of this is that the chains may be run simultaneously with parallel 
processing and yet still combined at the end provided that the burn-in trials 
have been removed and that each chain has reached adequate convergence.

\subsubsection{Convergence}

Convergence for each free parameter in each chain may be checked by computing 
the \citet{geweke:1992} statistic. This simple statistic compares a given
parameter's value at the beginning of the chain and at the end of the chain,
accounting for the variation due to parameter exploration. It is essentially
characterizes the number of sigmas difference these two points. In a converged
chain, we require that the \citet{geweke:1992} statistic $\lesssim 1$ and
certainly $\leq 3$.

Convergence is not generally expected with n-MAP since the problem is usually
under-constrained. Thus the application of the \citet{geweke:1992} statistic
is not employed for n-MAP results.

\subsubsection{Down-Sampling}

Sometimes a fit requires either a very long chain or multiple chains which are
combined. This can lead to a very large number of points in the final combined
chain, of order $10^6$-$10^7$. In general, this many points is excessive to
build reliable posteriors. One significant disadvantage of this is that
n-MAP must count the number of trials below/above various $\Theta_{ij}$ 
thresholds at every n-MAP MCMC realization. To expedite this process but
maintain the required level of precision, we evenly down-sample long MCMC chains 
from the light curve fits until there are only $10^5$ points remaining. Since 
the chains are evenly down-sampled, then the effective length of the chain is 
unaffected.

For the n-MAP plots, plotting $10^7$ points on a figure is excessive both
computationally and visually. Therefore, we down-sample any long n-MAP chains 
until we have $10^6$ points remaining. Once again, the effective lengths are
unaffected.


\section{Hypothetical Synthetic Systems}
\label{sec:testing}


\subsection{KOI-S01: A Moderate-Eccentricity Triple-System}
\label{sub:KOI-S01}

In order to demonstrate and test the MAP techniques discussed thus far, we will
present two hypothetical analyses in this section. Additional control tests
are also available in \S\ref{sub:KOI-S0P}, \ref{sub:KOI-S0G} \& 
\S\ref{sub:KOI-S0C}. 

For our first example, we consider a hypothetical three-planet system dubbed 
``KOI-S01'' for Kepler Object of Interest Synthetic 01. We use the same 
identification as that used for Kepler Objects of Interest because we envision 
that real KOI targets will be the most obvious application of our technique in 
the near future. The three planets (KOI-S01.01, KOI-S01.02 \& KOI-S01.03) are 
chosen to have orbital periods of $P_3 = 13.9342$\,d, $P_2 = 25.13654$\,d and 
$P_1 = 44.86254$\,d around a Solar-star. These were selected to provide at
least three transits for all three planets within the total time window of
the Q0, Q1 and Q2 Kepler data (127\,d). We also deliberately avoid mean motion
resonance to provide the plausible scenario that the planets follow a strictly 
linear ephemeris for the sake of simplicity (note that TTVs do not invalidate 
our technique and can be easily accounted for by allowing each transit to have 
an individual parameter for the time of transit minimum). The occultation depths 
are assumed to be negligible and the transit epoch for each planet is selected 
such that a) we obtain at least three transits for each planet b) no overlapping 
transits occur.

The eccentricity parameters were selected such 
that the Hill stability criterion \citep{gladman:1993} was satisfied: 

\begin{align}
\frac{ P_{\mathrm{outer}} }{ P_{\mathrm{inner}} } > \Bigg( 1 + \sqrt{\frac{8}{3} e_{\mathrm{inner}}^2 + 5.76 \Big(\frac{M_{\mathrm{inner}} + M_{\mathrm{outer}} }{M_*} \Big)^{2/3} } \Bigg)^{3/2}
\end{align}

As a result of this criterion, the eccentricities are therefore ``moderate'' and 
not large: $e_3 = 0.05$, $e_2 = 0.15$ and $e_1 = 0.08$. We also decided to 
enforce apsidal locking \citep{batygin:2009} to try to provide potentially 
realistic scenario. Since the transit probability is highest for planets near 
$\omega\sim90^{\circ}$ \citep{kane:2008}, we chose locking about this point: 
$\omega_3 = 92.23^{\circ}$, $\omega_2 = 89.42^{\circ}$ and 
$\omega_1 = 91.20^{\circ}$. Transit impact parameters were randomly generated
to be $b_3 = 0.276$, $b_2 = 0.055$ and $b_1 = 0.858$ and planetary radii were
arbitarily set to $R_3 = 0.45$\,$R_J$, $R_2 = 1.05$\,$R_J$ and 
$R_1 = 0.92$\,$R_J$. The properties of the KOI-S01 system are summarized in
Table~\ref{tab:KOI-S01}.

Quadratic limb darkening coefficients for the star were generated assuming
a Solar-like star and a \citet{kurucz:2006}-style atmosphere, giving 
$u_1=0.4277$ and $u_2=0.2522$. It should also be noted that the properties of
all three planets satisfy the criteria for MAP to work, as described in 
\S\ref{sub:psireliability} i.e. the $\Psi$-equation is accurate to better than 
99\% and the duration approximation is accurate to better than 99.9\%.

Synthetic data were generated to span the 127\,day window of the Q0, Q1 and
Q2 data of \emph{Kepler}. We chose to use short-cadence (58.84876\,s) data with
Gaussian noise of 250\,ppm (consistent with typical \emph{Kepler} noise for a
$V\sim12$ star e.g. \citealt{kippingbakos:2011b}) and random reference mean 
anomalies for the planets (although ensuring no mutual transits) yielding 
186,393 synthetic photometry points.

In a totally blind manner, one of us (DK) generated the synthetic data and
blindly passed it onto the other three (WD, JJ \& VM) who identified the
number of planets, the orbital periods and fitted them using an MCMC routine
coupled with the \citet{mandel:2002} algorithm. We then followed the steps
outlined in \S\ref{sub:algorithm}.

\begin{table*}
\caption{\emph{Physical properties of the hypothetical KOI-S01 system.
We assume $M_P\ll M_*$ during the calculation of the planetary
semi-major axis, thereby removing the need to assign planetary masses.}} 
\centering 
\begin{tabular}{c c c c c c c} 
\hline
Object & $M$ & $R$ & $P$ & $b$ & $e$ & $\omega$ \\ [0.5ex] 
\hline
KOI-S01.03 & - & 5.1\,$R_{\oplus}$ & 13.93\,d & 0.28 & 0.05 & $92.2^{\circ}$ \\
KOI-S01.02 & - & 11.7\,$R_{\oplus}$ & 25.14\,d & 0.05 & 0.15 & $89.4^{\circ}$ \\
KOI-S01.01 & - & 10.3\,$R_{\oplus}$ & 44.86\,d & 0.86 & 0.08 & $91.2^{\circ}$ \\ 
\hline
KOI-S01 & 1.00\,$M_{\odot}$ & 1.00\,$R_{\odot}$ & - & - & - & - \\ [1ex]
\hline\hline 
\end{tabular}
\label{tab:KOI-S01} 
\end{table*}

\subsubsection{Light curve fits}

\begin{figure}
\begin{center}
\includegraphics[width=8.4 cm]{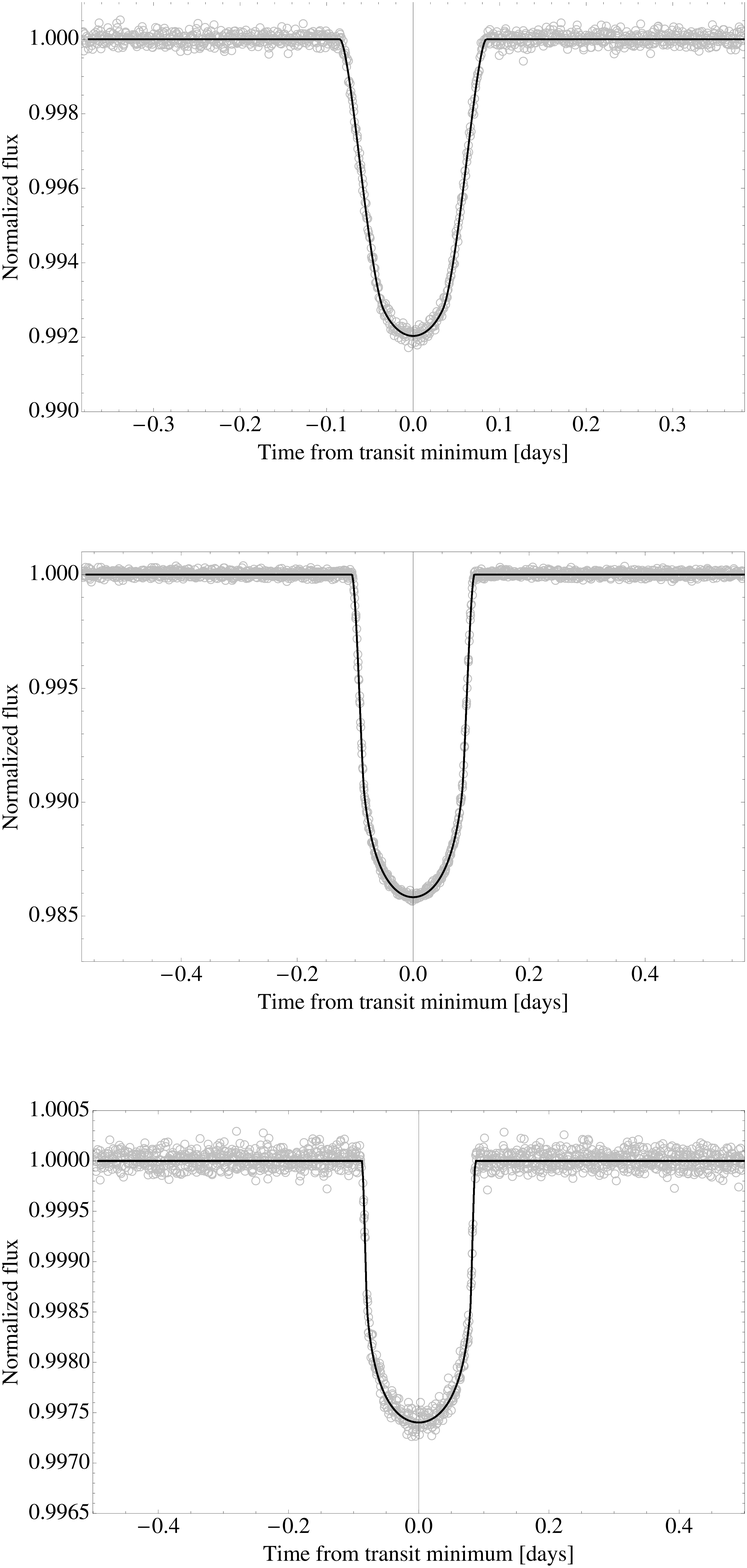}
\caption{\emph{
Folded light curves of KOI-S01.01 (top), KOI-S01.02 (middle) and 
KOI-S01.03 (bottom). The synthetic photometry is shown as gray circles, 
phase-binned with a bin size given by the number of transit epochs. The best-fit
transit model is shown as a continuous solid line in each case.}} 
\label{fig:lcplot1}
\end{center}
\end{figure}

\begin{figure*}
\begin{center}
\includegraphics[width=16.8 cm]{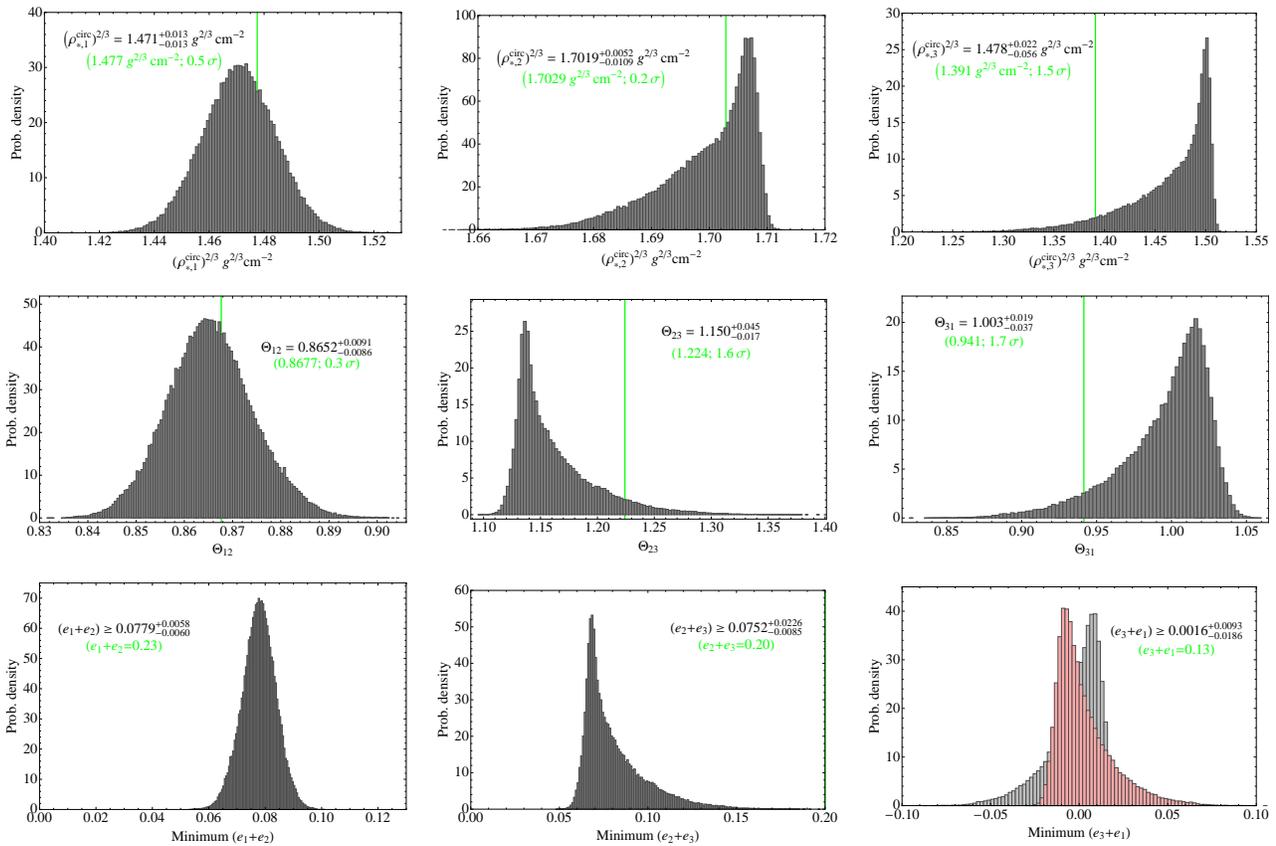}
\caption{\emph{Marginalized posteriors from the light curve fits (stage 1) only, 
for KOI-S01. In all figures, the green line marks the truth, with the numerical 
value provided in parentheses.
\textbf{Row 1:} Posteriors for $(\rho_{*,k}^{\mathrm{circ}})^{2/3}$. Fits assume 
a uniform prior in $(\rho_{*,k}^{\mathrm{circ}})^{2/3}$. Notice how even though 
the eccentricities in the system are not large, the derived stellar densities 
clearly indicate the presence of eccentricity in the system due to the 
non-overlapping posteriors.
\textbf{Row 2:} Posteriors for the ratios of the 
$(\rho_{*,k}^{\mathrm{circ}})^{2/3}$ terms.
\textbf{Row 3:} Posteriors for the ratios of the minimum pair-combined
eccentricities, computed using the approximations 
Equation~\ref{eqn:doubleA}\&\ref{eqn:doubleB} (i.e. a-MAP).}} 
\label{fig:histos1}
\end{center}
\end{figure*}

The light curve fits for all three planets are shown in Figure~\ref{fig:lcplot1} 
and the derived posteriors are presented in Figure~\ref{fig:histos1}. 
Diagnostics on the mixing and convergence of these MCMC fits are presented in 
Table~\ref{tab:diags_1}, all of which indicate reliable results.

As expected, the correct radii, transit epoch and orbital periods were
found in the blind-search. The derived stellar densities, assuming a circular
orbit, were found to be 
$(\rho_{*,1}^{\mathrm{circ}})^{2/3} = 1.471_{-0.013}^{+0.013}$\,g$^{2/3}$\,cm$^{-2}$,
$(\rho_{*,2}^{\mathrm{circ}})^{2/3} = 1.7019_{-0.0109}^{+0.0052}$\,g$^{2/3}$\,cm$^{-2}$ and
$(\rho_{*,3}^{\mathrm{circ}})^{2/3} = 1.478_{-0.056}^{+0.022}$\,g$^{2/3}$\,cm$^{-2}$,
which clearly deviate significantly from both a common value and the actual
stellar density of $(\rho_*)^{2/3} = 1.258$\,g$^{2/3}$\,cm$^{-2}$. Note that
we here quote the median of each marginalized distribution as the best-value
and the uncertainties come from the 34.15\% quantiles either side of the median.
This practice is continued for all results presented in this work.

\begin{table*}
\caption{\emph{MCMC diagnostics of fits for KOI-SO1. Diagnostics presented here
as discussed in \S\ref{sub:mcmcdiagnostics}.}} 
\centering 
\begin{tabular}{c c c c c c} 
\hline
Planet & \# of Accepted & Lowest eff. & Parameter w/ & Highest & Parameter w/ \\ [0.5ex] 
 & MCMC trials & length & lowest eff. len. & Geweke diag. & highest Geweke diag. \\ [0.5ex] 
\hline
KOI-S01.01 & $1\times(1.25\times10^5)$ & 6247 & $b$ & 0.028 & $\mathrm{OOT}_{-1}$ \\ 
KOI-S01.02 & $1\times(2.50\times10^5)$ & 1968 & $b$ & 0.0051 & $\mathrm{OOT}_{0}$ \\
KOI-S01.03 & $4\times(1.25\times10^5)$ & 1258 & $b$ & 0.032 & $\mathrm{OOT}_{+2}$ \\ 
\hline
n-MAP & $1\times(1.25\times10^6)$ & 1156 & $e_1$ & N/A & N/A \\ [1ex]
\hline\hline 
\end{tabular}
\label{tab:diags_1} 
\end{table*}

\subsubsection{Results using \MakeLowercase{a}-MAP}

The analytic approximations from \S\ref{sub:triples} may be used to provide 
lower bounds on combinations of $e_1$, $e_2$ and $e_3$ via 
Equation~\ref{eqn:triple}, for which we find:
$(e_1+e_2) \gtrsim 0.0779_{-0.0060}^{+0.0058}$ (true $e_1 + e_2 = 0.23$),
$(e_2+e_3) \gtrsim 0.0752_{-0.0085}^{+0.0226}$ (true $e_2 + e_3 = 0.20$) and
$(e_3+e_1) \gtrsim 0.0016_{-0.0186}^{+0.0093}$ (true $e_3 + e_1 = 0.13$). As
expected, all of these values are consistent with the true numbers. Further,
the significance of each combination being $>0$ is given by $13.0$-$\sigma$,
$8.8$-$\sigma$ and $0.1$-$\sigma$ respectively, thus indicating that the a-MAP 
method definitively shows that the system contains significant eccentricities. 
Given that $(e_3+e_1)$ is consistent with zero, one may correctly assert that
planet 2 is most likely to contain the majority of the net eccentricity.

\subsubsection{Results using \MakeLowercase{n}-MAP}

Using the n-MAP algorithm described in \S\ref{sub:algorithm}, we explored the 
full 6-dimensional permitted parameter space with $B = 1.25 \times 10^6$ MCMC 
trials. Jump sizes were selected to be 1\% for all 
terms (i.e. $\Delta e_k = 0.01$, $\Delta \omega_k = 0.01\times2\pi$\,rads). The 
starting point for the chain was randomly generated until a point with 
$\chi^2<10$ was located. Mixing was checked for as described in 
\S\ref{sub:mcmcdiagnostics} and the results are reported in 
Table~\ref{tab:diags_1}.

\begin{figure*}
\begin{center}
\includegraphics[width=16.8 cm]{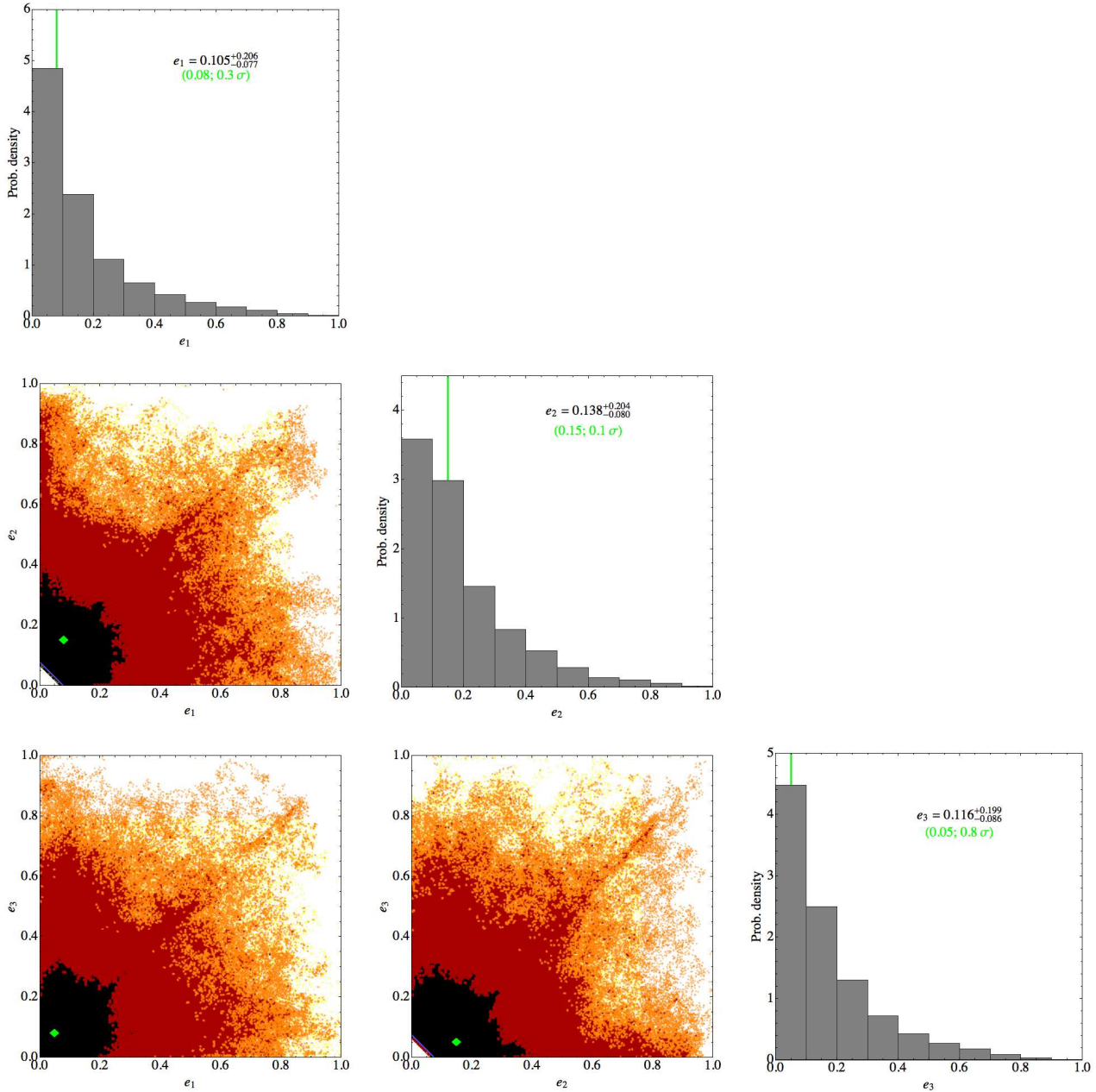}
\caption{\emph{
Results of n-MAP fits for KOI-S01 using uniform priors in $e_k$ and $\omega_k$. 
Green diamonds/lines mark the truth. Blue lines mark the median of the lower 
limits found using analytic MAP (shown in Figure~\ref{fig:histos1})
Black denotes 0-1\,$\sigma$, red denotes 1-2\,$\sigma$, orange 
denotes 2-3\,$\sigma$ and yellow denotes $>3$\,$\sigma$. White regions were
never vistted in any MCMC realizations and thus are highly improbable.}} 
\label{fig:nmap1}
\end{center}
\end{figure*}

The results are shown in Figure~\ref{fig:nmap1}, after down-sampling to one 
million trials. The joint posteriors clearly shows the minimum constraints on 
$(e_1+e_2)$ and $(e_2+e_3)$ (the white regions in the corner), as derived using 
a-MAP. We note that the diamonds shown in the joint posteriors of 
Figure~\ref{fig:nmap1} (which represent the true values), consistently lie in 
densely populated regions, supporting the validity of the MAP technique.The 1D 
marginalized posteriors yield $e_1 = 0.105_{-0.077}^{+0.206}$,
$e_2 = 0.138_{-0.080}^{+0.204}$ and $e_3 = 0.116_{-0.199}^{+0.086}$, all consistent 
with the truth to within 1$\sigma$.

With the a-MAP results, it was shown that one may reasonably deduce that
planet 2 is most likely to sustain the highest eccentricity in the system.
n-MAP agrees with this conclusion since $e_2$ has the highest median of all
three marginalized posteriors. However, the significance of the orbital
eccentricities for all three planets appears marginal. The significances of
$e_k>0$ for $k=1,2,3$ are 1.4-$\sigma$, 1.7-$\sigma$ and 1.4-$\sigma$ 
respectively, which a far cry from the $13.0$-$\sigma$ level detections which
were achieved by a-MAP (on the same data). This suggests the following paradigm 
about the results from MAP (both a-MAP \& n-MAP): \emph{MAP is more sensitive to 
pair-combined eccentricities than individual eccentricities}.

If we had not used n-MAP but assumed uniform priors in $e_k$, the probability
that $e_1<0.3$ (which is a useful rough limit for a habitable world) would be 
30\% and the probability that $e_1>0.3$ would be 70\%. Thus it would be 
$\sim0.4$ times more likely that the orbit was $e_1<0.3$ than otherwise. The 
same is of course true for $e_2$ and $e_3$. Using n-MAP these odds ratios become 
2.4, 1.2 and 2.7 for $e_1<0.3$, $e_2<0.3$ and $e_3<0.3$ respectively, 
demonstrating the extra information we have gained from using n-MAP.


\subsection{KOI-S02: 61-Virginis Analog System}
\label{sub:KOI-S02}

\subsubsection{Setup}

As an additional test, we decided to look at a genuine multiple planet system
with well-characterized eccentricities. In order to satisfy this requirement
and additionally locate a system with $\geq3$ planets, we must draw upon planets
found through radial velocity, rather than the transit technique. This is
because RV planets have much better orbital solutions than the few multiple
systems found by \emph{Kepler} so far. As the systems are RV planets and not
transiting, we must generate synthetic photometry for them, in a similar way
as to was done for KOI-S01.

From the 12 systems which satisfy our criteria at the time of writing 
(according to www.exoplanets.org), we selected the 61-Virginis system. 
61-Virginis has orbital periods short enough to be detected within the first 18 
months of operation of \emph{Kepler} and posseses the highest orbital 
eccentricity components from such systems. Despite the eccentricities
being the largest found from those available systems, all three planets in the
61-Virginis system satisfy the criteria for MAP to work, as described in 
\S\ref{sub:psireliability} i.e. the $\Psi$-equation is accurate to better than 
99\% and the duration approximation is accurate to better than 99.9\%.

The relevant properties of the 61-Virginis are provided in 
Table~\ref{tab:KOI-S02} and taken from \citet{vogt:2010}. We dub our 
hypothetical system as KOI-S02, to stress the fact that this is a hypothetical 
analysis and not a genuine study of 61-Virginis.

\begin{table*}
\caption{\emph{Physical properties of the hypothetical KOI-S02 system, whose 
properties are assumed to be the same as that as the 61-Virginis system from 
\citet{vogt:2010}. The radii and impact parameters have been arbitarily 
assigned, since the system is not known to have any transiting members.}} 
\centering 
\begin{tabular}{c c c c c c c c} 
\hline
Object & Object Analog & $M$ & $R$ & $P$ & $b$ & $e$ & $\omega$ \\ [0.5ex] 
\hline
KOI-S02.03 & 61-Vir b & 5.1\,$M_{\oplus}$ & 1.6\,$R_{\oplus}$ & 4.215\,d & 0.15 & 0.10 & $110^{\circ}$ \\
KOI-S02.02 & 61-Vir c & 10.5\,$M_{\oplus}$ & 3.3\,$R_{\oplus}$ & 38.02\,d & 0.40 & 0.14 & $340^{\circ}$ \\
KOI-S02.01 & 61-Vir d & 22.9\,$M_{\oplus}$ & 4.3\,$R_{\oplus}$ & 123.0\,d & 0.75 & 0.35 & $310^{\circ}$ \\ 
\hline
KOI-S02 & 61-Vir & 0.94\,$M_{\odot}$ & $0.979$\,$R_{\odot}$ & - & - & - & - \\ [1ex]
\hline\hline 
\end{tabular}
\label{tab:KOI-S02} 
\end{table*}

In order to have guarantee that we have measured three transits of all three 
objects (assuming all three indeed transit), we would require $4P_d$ days of
continuous photometry. To mimic this, we consider 1.5\,years of \emph{Kepler}
short-cadence photometry. We estimated our noise based upon a simple 
calculation. The $V=11.4$ star TrES-2 has been observed in \emph{Kepler} 
short-cadence mode to have an RMS noise of 237\,ppm per minute 
\citep{kippingbakos:2011b}. As 61-Virginis is $V=4.74$, it would never have been
observed by \emph{Kepler} because it is too bright. Nevertheless, consider the
star was at the bright limit of \emph{Kepler}'s range, namely $V\simeq9$, then
the star would be at the floor-limit of RMS precision. This should be around
25\% lower than the RMS for a $V=11.4$ star. Accordingly, we assigned an RMS
precision of 178\,ppm per minute for this synthetic data set.

As the planets do not transit, we had to assign planetary radii. We made the
simple assumption that those planets of mass $M\geq10$\,$M_{\oplus}$ 
(61-Vir c \& d) were ice/gas giants similar in composition to Neptune. We 
accordingly computed their radius assuming the average bulk density was equal to 
$\rho_{\mathrm{Neptune}}=1.628$\,g\,cm$^{-3}$. For planets of mass (61-Vir b), 
we assumed the simple mass-radius scaling law of a terrestrial planet 
$R\sim M^{0.27}$ \citep{valencia:2006} in Earth-mass and -radii units.

Limb darkening was assumed to be the same as with KOI-S01 for simplicity. The 
transit impact parameters were arbitrarily chosen to be $b_b = 0.15$, 
$b_c = 0.40$ and $b_d = 0.75$.

\subsubsection{Light curve fits}

\begin{figure}
\begin{center}
\includegraphics[width=8.4 cm]{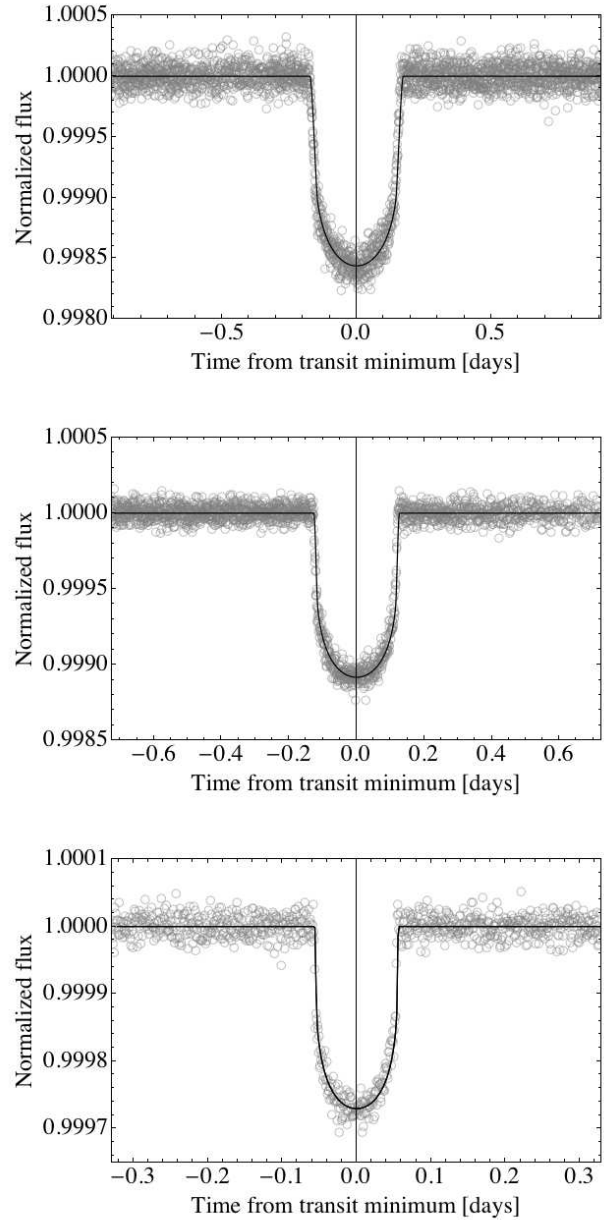}
\caption{\emph{
Folded light curves of KOI-S02.01 (top), KOI-S02.02 (middle) and 
KOI-S02.03 (bottom). The synthetic photometry is shown as gray circles, 
phase-binned with a bin size given by the number of transit epochs. The best-fit
transit model is shown as a continuous solid line in each case.}} 
\label{fig:lcplot2}
\end{center}
\end{figure}

\begin{figure*}
\begin{center}
\includegraphics[width=16.8 cm]{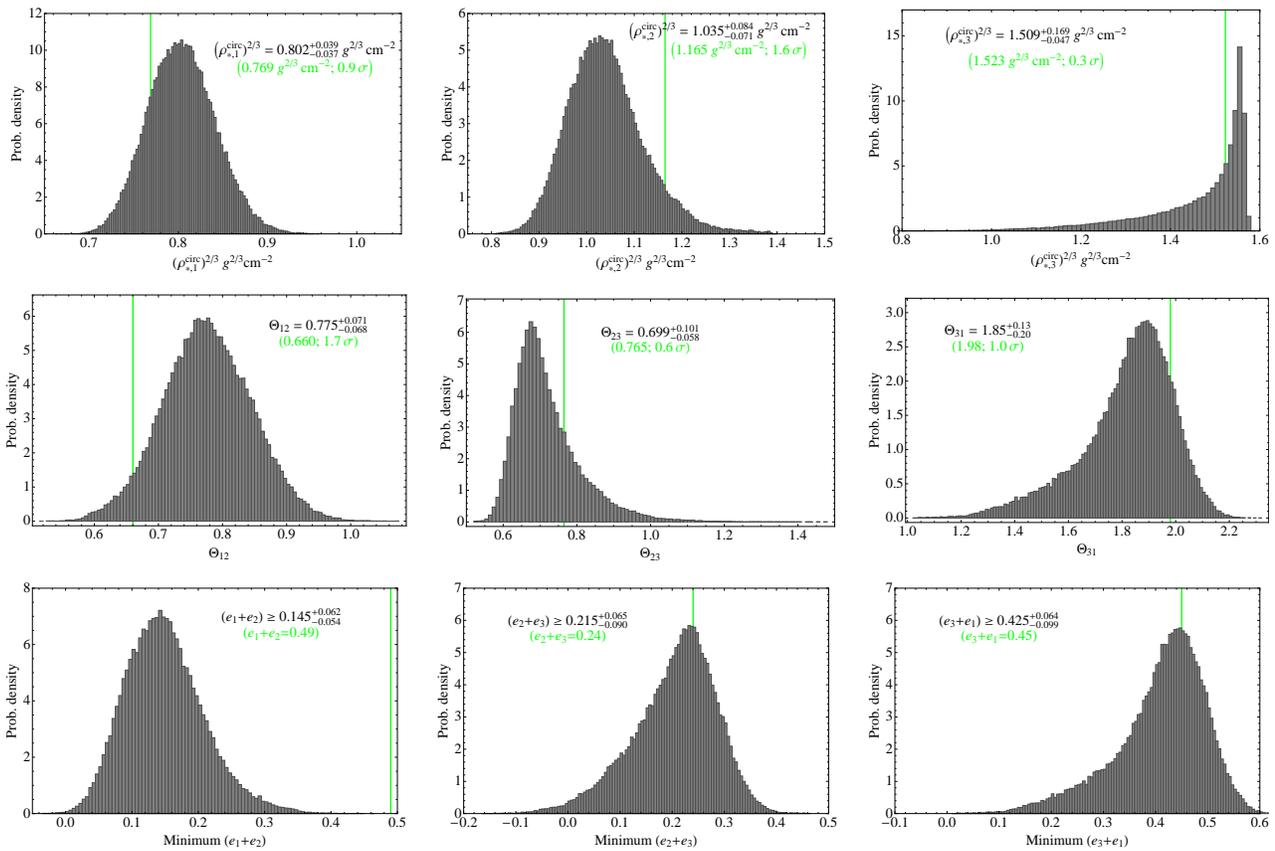}
\caption{\emph{Marginalized posteriors from the light curve fits (stage 1) only, 
for KOI-S02. In all figures, the green line marks the truth, with the numerical 
value provided in parentheses.
\textbf{Row 1:} Posteriors for $(\rho_{*,k}^{\mathrm{circ}})^{2/3}$. Fits assume 
a uniform prior in $(\rho_{*,k}^{\mathrm{circ}})^{2/3}$.
\textbf{Row 2:} Posteriors for the ratios of the 
$(\rho_{*,k}^{\mathrm{circ}})^{2/3}$ terms.
\textbf{Row 3:} Posteriors for the ratios of the minimum pair-combined
eccentricities, computed using the approximatations 
Equation~\ref{eqn:doubleA}\&\ref{eqn:doubleB} (i.e. a-MAP).}} 
\label{fig:histos2}
\end{center}
\end{figure*}

The light curve fits for all three planets are shown in Figure~\ref{fig:lcplot2} 
and the derived posteriors are presented in Figure~\ref{fig:histos2}. 
Diagnostics on the mixing and convergence of these MCMC fits are presented in 
Table~\ref{tab:diags_2}, all of which indicate reliable results.

As expected, the correct radii, transit epoch and orbital periods were
recovered. The derived stellar densities, assuming a circular
orbit, were found to be 
$(\rho_{*,1}^{\mathrm{circ}})^{2/3} = 0.802_{-0.037}^{+0.039}$\,g$^{2/3}$\,cm$^{-2}$,
$(\rho_{*,2}^{\mathrm{circ}})^{2/3} = 1.035_{-0.071}^{+0.084}$\,g$^{2/3}$\,cm$^{-2}$ and
$(\rho_{*,3}^{\mathrm{circ}})^{2/3} = 1.509_{-0.047}^{+0.169}$\,g$^{2/3}$\,cm$^{-2}$,
which clearly deviate significantly from both a common value and the actual
stellar density of $(\rho_*)^{2/3} = 1.260$\,g$^{2/3}$\,cm$^{-2}$.

\begin{table*}
\caption{\emph{MCMC diagnostics of fits for KOI-SO2. Diagnostics presented here
as discussed in \S\ref{sub:mcmcdiagnostics}.}} 
\centering 
\begin{tabular}{c c c c c c} 
\hline
Planet & \# of Accepted & Lowest eff. & Parameter w/ & Highest & Parameter w/ \\ [0.5ex] 
 & MCMC trials & length & lowest eff. len. & Geweke diag. & highest Geweke diag. \\ [0.5ex] 
\hline
KOI-S02.01 & $1\times(1.25\times10^5)$ & 4808 & $b$ & 0.0048 & $\mathrm{OOT}_{0}$ \\ 
KOI-S02.02 & $1\times(1.25\times10^5)$ & 2232 & $b$ & 0.0074 & $\mathrm{OOT}_{0}$ \\
KOI-S02.03 & $4\times(1.25\times10^5)$ & 1377 & $b$ & 0.0027 & $\mathrm{OOT}_{+2}$ \\ 
\hline
n-MAP & $1\times(1.25\times10^6)$ & 1247 & $e_3$ & N/A & N/A \\ [1ex]
\hline\hline 
\end{tabular}
\label{tab:diags_2} 
\end{table*}

\subsubsection{Results using \MakeLowercase{a}-MAP}

The analytic approximations from \S\ref{sub:triples} may be used to provide 
lower bounds on combinations of $e_1$, $e_2$ and $e_3$ via 
Equation~\ref{eqn:triple}, for which we find:
$(e_1+e_2) \gtrsim 0.145_{-0.054}^{+0.062}$ (true $e_1 + e_2 = 0.49$),
$(e_2+e_3) \gtrsim 0.215_{-0.090}^{+0.065}$ (true $e_2 + e_3 = 0.24$) and
$(e_3+e_1) \gtrsim 0.425_{-0.099}^{+0.064}$ (true $e_3 + e_1 = 0.45$). As
expected, all of these values are consistent with the true numbers. Further,
the significance of each combination being $>0$ is given by $2.7$-$\sigma$,
$2.4$-$\sigma$ and $4.3$-$\sigma$ respectively, therefore the a-MAP 
method strongly indicates that the system contains non-zero eccentricities.

\subsubsection{Results using \MakeLowercase{n}-MAP}

Using the n-MAP algorithm described in \S\ref{sub:algorithm}, we explored the 
full 6-dimensional permitted parameter space with $B = 1.25 \times 10^6$ MCMC 
trials. Jump sizes were selected to be 1\% for all 
terms (i.e. $\Delta e_k = 0.01$, $\Delta \omega_k = 0.01\times2\pi$\,rads). The 
starting point for the chain was randomly generated until a point with 
$\chi^2<10$ was located. Mixing was checked for as described in 
\S\ref{sub:mcmcdiagnostics} and the results are reported in 
Table~\ref{tab:diags_2}.

\begin{figure*}
\begin{center}
\includegraphics[width=16.8 cm]{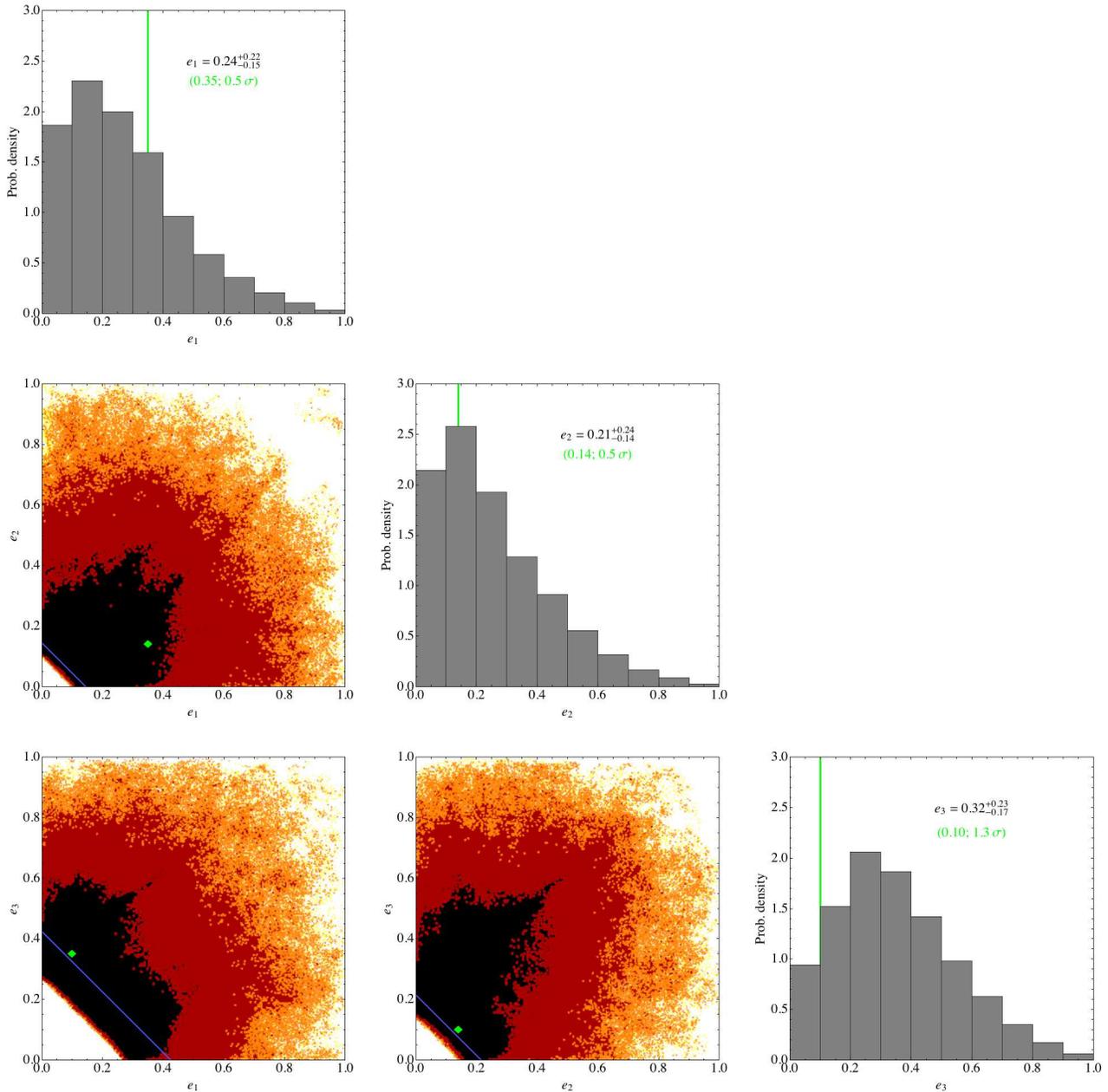}
\caption{\emph{
Results of n-MAP fits for KOI-S02 using uniform priors in $e_k$ and $\omega_k$. 
Green diamonds/lines mark the truth. Blue lines mark the median of the lower 
limits found using analytic MAP (shown in Figure~\ref{fig:histos2})
Black denotes 0-1\,$\sigma$, red denotes 1-2\,$\sigma$, orange 
denotes 2-3\,$\sigma$ and yellow denotes $>3$\,$\sigma$. White regions were
never vistted in any MCMC realizations and thus are highly improbable.
$e_1$ vs $e_3$ plot has a displaced blue line due to the large uncertainty
on this constraint, as evident in Figure~\ref{fig:histos2}.}} 
\label{fig:nmap2}
\end{center}
\end{figure*}

The results are shown in Figure~\ref{fig:nmap2}, after down-sampling to one 
million trials. The joint posteriors clearly shows the minimum constraints on 
all three pair-combinations (the white regions in the corner), as derived using 
a-MAP. We note that the diamonds shown in the joint posteriors of 
Figure~\ref{fig:nmap2} (which represent the true values), 
consistently lie in densely populated regions, supporting the validity of the 
MAP technique. The 1D marginalized posteriors yield 
$e_1 = 0.24_{-0.15}^{+0.22}$, $e_2 = 0.21_{-0.14}^{+0.24}$ and 
$e_3 = 0.32_{-0.17}^{+0.23}$, all consistent with the truth to within 1-$\sigma$
except $e_3$, which is marginally overestimated by 1.3-$\sigma$. Note that the 
mode tends to overestimate the eccentricity because eccentricity is a positive 
definite quantity \citep{lucy:1971}. These results, as with KOI-S01 
(\S\ref{sub:KOI-S01}) support the validity of the MAP technique.

In the earlier case of KOI-S01 (\S\ref{sub:KOI-S01}), Figure~\ref{fig:nmap1} 
showed that when the median of $(e_i+e_j)_{\mathrm{min}}$ posterior (derived 
using a-MAP) was plotted along with the 2D-marginalized posteriors of $e_i$ 
versus $e_j$ (derived using n-MAP), there existed excellent agreement on the 
minimum bound on the pair-combination. In contrast, the KOI-S02 simulation 
reveals one interesting exception to this pattern, specifically for 
$(e_3+e_1)_{\mathrm{min}}$, as seen in Figure~\ref{fig:namp2}. Here, the median 
of the $(e_3+e_1)_{\mathrm{min}}$ posterior (derived using a-MAP) yields 
$(e_3+e_1)\geq0.425$ whereas visual inpsection of the n-MAP results suggests 
$(e_3+e_1)\geq0.275$. Including the a-MAP uncertainties reveals 
$(e_3+e_1)\geq0.425_{-0.099}^{+0.064}$ and thus is $1.5$-$\sigma$ deviant from 
the n-MAP result. Whilst this is not statistically significant (and in fact both 
the a-MAP and the n-MAP limits are consistent with the truth of 
$(e_3+e_1)=0.45$), this is still a large departure relative to the other cases, 
and may lead the reader to question the origin of this discrepancy.

The reason for the discrepancy can be understood in terms of the approximations
made in the original derivation of a-MAP. In \S\ref{sub:doubles}, the derivation
of a-MAP includes a step where we assume $e_i\ll1$, allowing us to execute a
first-order series expansion. Therefore, one can see that the higher
the eccentricity terms, the more a-MAP will depreciate as a reliable 
approximation. In the KOI-S02 case, the true values of the eccentricity
pair-combinations are $(e_1+e_2)=0.23$, $(e_2+e_3)=0.20$ and $(e_3+e_1)=0.45$.
Consequently, it is clear that $(e_3+e_1)$ is exposed to the highest 
eccentricity and thus one should expect that a-MAP would be the least reliable 
for this case. Indeed, this is exactly what is seen, explaining the anomalous 
behaviour discussed in the previous paragraph. Despite the large pair combined
eccentricity of 0.45, it is reassuring that a-MAP still performs quite well,
landing to within $1.5$-$\sigma$ of the more sophisticated n-MAP result.

\section{Discussion \& Conclusions}
\label{sec:discussion}

\subsection{Recurring Patterns in MAP}
\label{sub:patterns}

In the 2D posteriors shown in the example systems KOI-SOG, KOI-SOC, KOI-S01
and KOI-S02, it is possible to visually detect some recurring patterns. In this
subsection, we will discuss the reason for these patterns.

The most obvious pattern is that the plots tend to exhibit higher densities
for $e_x \simeq e_y$. This results in a diagonal high density region extending
from the bottom-left to the top-right, getting narrower as it goes up. Why
does this happen?

Consider two planets on circular orbits. The light curve derived stellar
densities will be identical meaning MAP will find that a system with two
circular orbits to be a highly compatible solution. However, it is not the only
compatible solution. Consider a MAP realization with $e_1 = 0$ and $e_2 = 0.1$.
In this case, a rather broad range of $\omega_2$ values can reproduce a stellar
density which is still equivalent (or within error) to 
$\rho_{*,1}^{\mathrm{circ}}$. As $e_2$ increases to higher eccentricities
(i.e. as we move along one of the axes), the range of $\omega_2$ values which 
can still reproduce such a result diminishes. Consequently, the integrated 
probability density over all $\omega_2$ values is small and thus the region 
becomes a low-probability sector.

Now consider MAP realizations where both $e_1$ and $e_2$ depart from zero.
In this case, a situation where $e_2 = 0.9$ and $e_1 = 0.1$ is not grossly
different from the case just described where $e_2$ is very large and $e_1$
is zero. Thus these regions tend to be low probability. However, if
$e_1 \simeq e_2$ and both values are large then the range of $\omega_1$
and $\omega_2$ values which can reproduce $\rho_{*,1}^{\mathrm{circ}} \simeq 
\rho_{*,2}^{\mathrm{circ}}$ is broader. Thus the integrated probability density
becomes larger too. This can be seen be simple inspection of 
Equation~\ref{eqn:rhoratio}. For this reason, a common pattern seen in MAP
is the tail extending along the $e_x = e_y$ axis.

\subsection{MAP in Combination with Other Observables}
\label{sub:MultiMAP}

Multibody Asterodensity Profiling (MAP) can provide constraints on the orbital
eccentricity of transiting planets in multiple planet systems without the need
for radial velocities, occultations or transit timing variations. However,
there may some cases where these observables are available. In such a case,
MAP can be combined with these other observables to further refine the
constraints on the orbital eccentricities. Our n-MAP technique utilizes an
MCMC routine to explore the parameter space of possible orbital eccentricities.
This MCMC works by assigning a $\chi^2$ merit-function to each point and then
proceeding via the Metropolis-Hastings rule. Using n-MAP alone, the merit
function is:

\begin{align}
\chi^2 &= \chi_{\mathrm{MAP}}^2
\end{align}

where $\chi_{\mathrm{MAP}}^2$ is given in Equation~\ref{eqn:chi2}. If
additional observables are available, such as TTVs or RVs then one may simply
append the associated merit functions:

\begin{align}
\chi^2 &= \chi_{\mathrm{MAP}}^2 + \chi_{\mathrm{RV}}^2 + \chi_{\mathrm{TTV}}^2 + ...
\end{align}

Once appended, the n-MAP routine is simply executed as before. In our
next paper, we will provide an analysis of this technique on a real transiting
planet system featuring both radial velocities and transit timing variations.

\subsection{Differences to the \citet{moorhead:2011} Technique}

A direct comparison to the method of M11 is not fair because the techniques 
operate under different conditions and assumptions. Despite this, we will
here outline a few differences between the two methods. One advantage of the M11 
method is that it works for all transiting planets whereas MAP requires $>1$ 
transiting planet in a given system. One advantage of MAP over the M11 technique 
is that MAP is highly model independent\footnote{There is a very weak model
dependency in MAP via the adopted limb darkening law, but this can also be
fitted for with sufficient signal to noise}, assuming only the planets orbit the 
same star whereas M11 require a value for the stellar radius determined
via the more model sensitive route of stellar evolution.

In principle, MAP exploits more information in the light curve than that of
the M11 technique. M11 compare the observed transit duration to the maximum
theoretical value for a circular orbit; in other words M11 make use of one
metric for the duration. Typically, this metric is the transit duration defined
as the time for the planet to move from its centre crossing the stellar limb
to exitting under the same condition, $\tilde{T}$, since this is independent
of the derived planetary radius. In contrast, MAP uses both $\tilde{T}$ and
the ingress duration, $T_{12}$, to derive the light curve derived stellar 
density assuming a circular orbit, $\rho_{*}^{\mathrm{circ}}$. Thus, it can be
appreciated that MAP uses the same information as M11 plus some extra 
information. In the limit of ignoring this ingress information, the fundamental
transit information used by both techniques would be identical.

As a result of M11 negating the ingress duration information, the impact 
parameter is unresolved. For this reason, M11 adopt the conservative assumption
that the limiting case is for $b=0$. \citet{ford:2008} alternatively discuss how 
a prior in $b$ could be adopted by assuming an isotropic distribution in
orbital inclination. In contrast, MAP takes the distribution in $b$ from the
data itself, essentially characterized by the ingress duration. Due to the 
$b=0$ conservative assumption made by M11, only transit durations longer than
this limiting case can ever be detected. Unlike MAP, this limits the method to 
detecting eccentric planet transiting near apocentre, the slowest part of the 
orbit, and also the least likely geometric configuration to detect a transiting 
planet in \citep{kane:2008} (this is also pointed out in 
\citealt{tingley:2005}). This bias, discussed in M11, requires de-biasing any 
eccentricity statistics deduced and of course reduces the overall sample size 
since only a subset of eccentric planets are detected. For this reason, we 
anticipate MAP would provide a more powerful diagnostic of the statistics of 
eccentric planets, but of course is only applicable in multiple systems.

\subsection{Single-body Asterodensity Profiling (SAP)}
\label{sub:SAP}

MAP makes no assumption about the properties of the parent star. For stars
with poor characterization, this is an advantage since the stellar properties
are frequently subject to unknown systematic uncertainties. However, in some
cases the stellar properties are well-characterized and this is information
which MAP ignores. An example of this is a star which has been studied with
asteroseismology, leading to a highly precise determination of $\rho_*$.

Single-body Asterodensity Profiling (SAP) is the logical extension of MAP
which can include this information. In this work, we argue in favour of not
using SAP due to the frequently unreliable stellar parameters and the strong
model dependency of those derived results, which is in stark contrast to the
MAP technique. However, for sake of completion we will discuss here a possible 
implementation of SAP.

In each MCMC trial of the n-MAP algorithm, we create a \{$e_k,\omega_k$\} trial 
vector. This may be used to construct a $\{\Psi_k\}$ vector. For a single 
planet, this would be simply be a 1-dimensional vector: $\{\Psi_1\}$. This value 
may be used to infer the true stellar density based upon the derived value of 
$\rho_{*,1}^{\mathrm{circ}}$ via:

\begin{align}
\rho_{*,1}^{\mathrm{SAP}} &= \rho_{*,1}^{\mathrm{circ}}/\Psi_1 
\end{align}

For each MCMC trial, we can use the same value of $\rho_{*,1}^{\mathrm{circ}}$,
namely the median of the light curve derived posterior. We now simply
define $\chi^2_{\mathrm{SAP}}$ as the number of standard deviations between
this trial value of $\rho_{*,1}^{\mathrm{MAP}}$ and the empirically determined
value of $\rho_*$, say $\rho_{*}^{\mathrm{seismology}}$:

\begin{align}
\chi_{\mathrm{SAP},k}^2 &= \frac{[\rho_{*,k}^{\mathrm{SAP}} - \rho_{*}^{\mathrm{seismology}}]^2}{[\sigma(\rho_{*,k}^{\mathrm{SAP}})]^2 + [\sigma(\rho_{*}^{\mathrm{seismology}})]^2}
\end{align}

where we have replaced the 1 subscript (for planet ``1'') with $k$ to
make it a general equation. Clearly for $n$ planets we have $n$ contributions
to the total $\chi^2$ function from SAP:

\begin{align}
\chi_{\mathrm{SAP}}^2 &= \sum_{k=1}^n \chi_{\mathrm{SAP},k}^2 
\end{align}

This may be combined with the MAP, RV or TTV merit functions as desired to
produce finer constraints on the orbital eccentricity.

\subsection{Overview}
\label{sub:overview}

In this work, we have presented a new method to photometrically constrain
the orbital eccentricities of transiting planets. The method is only applicable
to multiple transiting planet systems and relies on the key assumption that all 
of the transiting planets orbit the same star. The new method works by comparing
the light curve derived stellar density between each planet and thus is dubbed
``Multibody Asterodensity Profiling'' (MAP). MAP requires no prior information
on the star's properties and thus is highly robust against systematic
uncertainties.

MAP constitutes a new observable for which the likelihood of a given orbital
configuration can be computed in a $\chi^2$-sense. Thus, MAP be be combined with
other pieces of information about the eccentricity of the system e.g.
transit timing variations, radial velocities, occultations. In this work,
we have adopted uniform priors for the orbital eccentricities but more realistic
priors based upon dynamics or planet formation can also be invoked (see 
\S\ref{sub:directpriors}).

In its simplest form, MAP can be applied by employing some simple analytic 
expressions (a-MAP) to deduce the combined eccentricities of two planets (e.g. 
Equation~\ref{eqn:doubleA}). In this sense, MAP exhibits impressive sensitivity
to a system with even a moderate to low eccentricity. For example, for a
synthetic system with one planet with $e=0.15$ (KOI-S01, see 
\S\ref{sub:KOI-S01}), Q0-Q2 Kepler data can infer a significant eccentricity at 
the 13-$\sigma$ level.

To determine the individual eccentricities require the use of numerical methods,
or n-MAP. n-MAP is shown to recover the same constraints for the pair-combined 
eccentricities as a-MAP does, but additionally provides individual constraints
as well as a fuller picture of the inter-relationships between the various
eccentricity terms. However, we find MAP is more sensitive to the
minimum pair-combined eccentricities than individual terms. Nevertheless,
an empricial determination of the posterior for each term is always derivable.
The highly model independent nature of MAP means that these posteriors will
narrow ad-infinitum as more data and signal-to-noise is accumulated. Further,
the application of MAP to dozens of systems raises the potential of discerning
truly unbiased statistics on the eccentricity distribution of planets in
multiple systems.

Ultimately, MAP has the potential to characterize the eccentricity of the first
truly habitable Earth-like planet, which could be found by \emph{Kepler}. In
such a case, the Earth-mass planet will likely be too challenging to detect
with radial velocities but MAP will be an ever-present tool provided the
photometric time series is of reasonable quality and the system has more
than one transiting planet (true of $\sim$50\% of all transiting planet 
candidates found by \emph{Kepler}). In conclusion, MAP offers a method
to diagnose both interesting dynamical systems and potentially interesting
astrobiological targets too.

The n-MAP algorithm is available as a Fortran 90 script upon request.


\section*{Acknowledgments}

DMK is supported by the NASA Carl Sagan fellowship scheme. Thanks to G. Bakos
for stimulating conversations on this research. WD, JJ and VM would like to 
thank G. Tinetti for her support and advice during this project. We 
would like to offer special thanks to our referee, Brandon Tingley, for his 
insightful comments and suggestions which greatly improved the manuscript.



\appendix

\section{Further Hypothetical Synthetic Systems}
\label{sec:testing2}

In this appendix, we will present several additional test systems on which we
implemented the MAP technique. The systems are control cases uses to ensure
we retrieve the correct null results.


\subsection{KOI-S0P: Triple System Fit Driven by Uniform Priors Only}
\label{sub:KOI-S0P}

\subsubsection{Setup}

\begin{figure*}
\begin{center}
\includegraphics[width=16.8 cm]{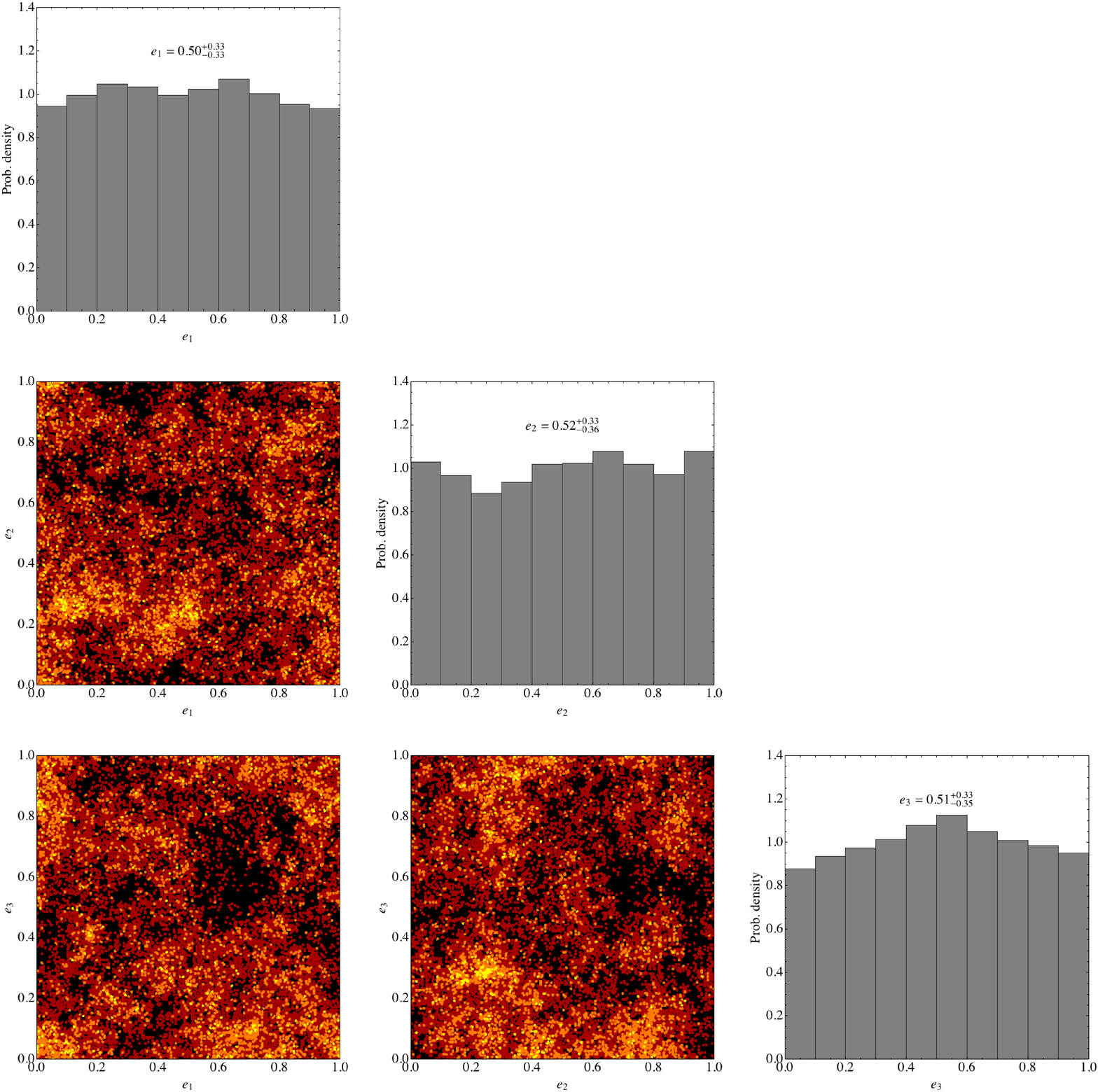}
\caption{\emph{
Results of n-MAP fits using no data input and uniform priors in $e_k$
and $\omega_k$. As expected, uniform posteriors are reproduced in all cases.
Black denotes 0-1\,$\sigma$, red denotes 1-2\,$\sigma$, orange 
denotes 2-3\,$\sigma$ and yellow denotes $>3$\,$\sigma$.}} 
\label{fig:nmapp}
\end{center}
\end{figure*}

We repeated our analysis of KOI-S01 (see \S\ref{sub:KOI-S01} for details) but in 
the second MCMC chain instructed the algorithm to accept all jumps, regardless 
of the likelihood. This instruction causes the algorithm to produce a result 
which purely reproduces the initial priors and thus is a useful of way of 
visualizing what an n-MAP result appears like in the absence of any 
observational constraints. We dub this system KOI-S0P (for ``priors''). The 
resulting n-MAP posteriors are shown in Figure~\ref{fig:nmapp}, which clearly 
reproduce the expected behaviour of uniform priors. 

\subsection{KOI:S0G: Triple System with Normally Distributed 
$(\rho_{*,k}^{\mathrm{circ}})^{2/3}$}
\label{sub:KOI-S0G}

\subsubsection{Setup}

\begin{figure*}
\begin{center}
\includegraphics[width=16.8 cm]{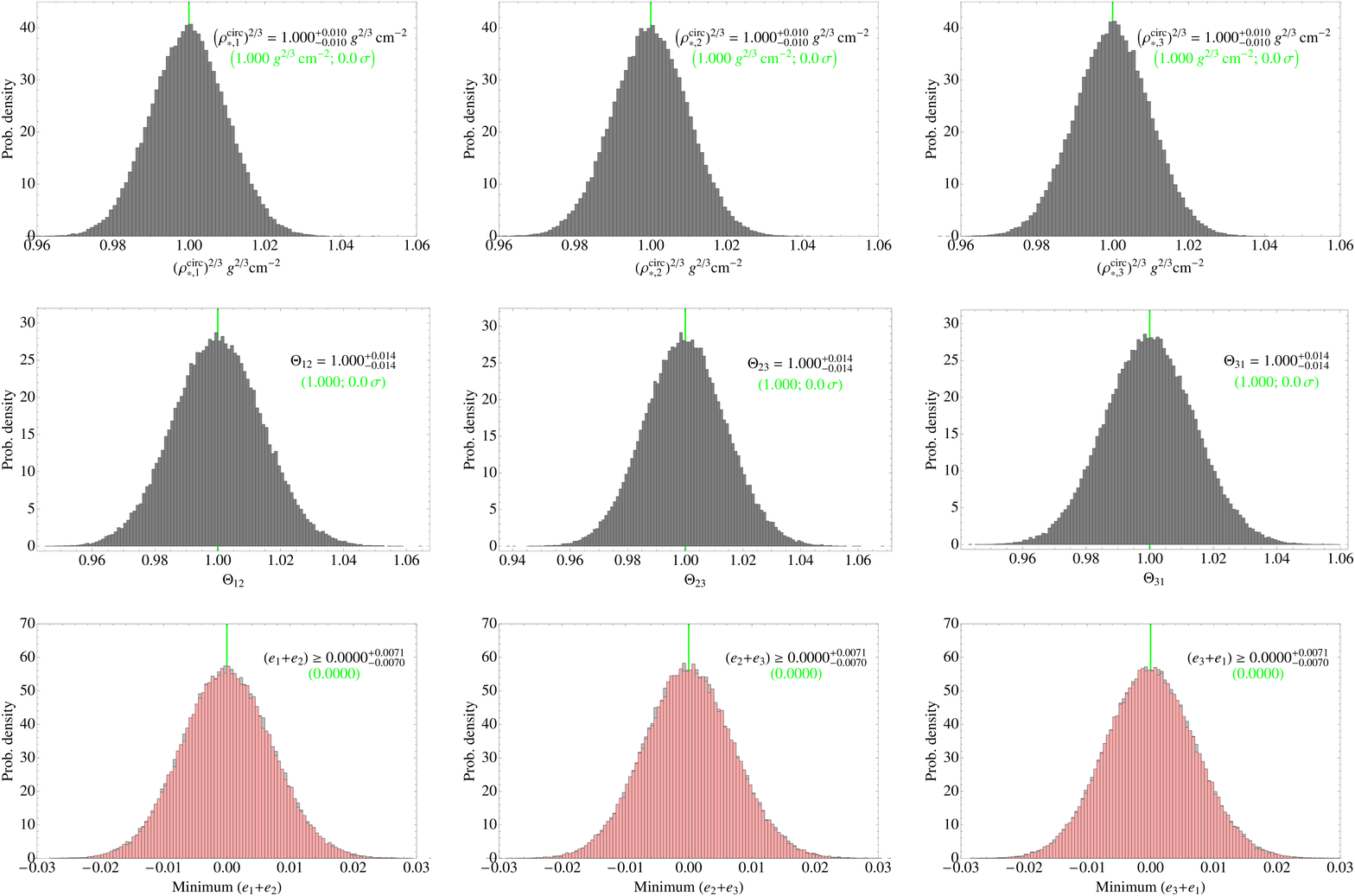}
\caption{\emph{Marginalized posteriors based upon synthetic 
$(\rho_{*,k}^{\mathrm{circ}})^{2/3}$ distributions, chosen to be perfectly 
Gaussian (KOI-S0G). In all figures, the green line vertical line marks the 
truth, with the numerical value provided in parentheses.
\textbf{Row 1:} Posteriors for $(\rho_{*,k}^{\mathrm{circ}})^{2/3}$. These have 
been synthetically generated to be perfectly Gaussian.
\textbf{Row 2:} Posteriors for the ratios of the 
$(\rho_{*,k}^{\mathrm{circ}})^{2/3}$ terms.
\textbf{Row 3:} Posteriors for the ratios of the minimum pair-combined
eccentricities, computed using the approximate expressions 
Equation~\ref{eqn:doubleA}\&\ref{eqn:doubleB}.
}} 
\label{fig:histosg}
\end{center}
\end{figure*}

For our second control system, we generated three independent distributions for 
$(\rho_{*,k}^{\mathrm{circ}})^{2/3}$ of $10^5$ points each. Each 
$(\rho_{*,k}^{\mathrm{circ}})^{2/3}$ realization is generated with random 
numbers selected from a normal distribution centred about $\mu=1$ with standard 
deviation $\sigma=0.01$. This allows us to see the isolated behaviour of MAP 
without any concern for potential biases from the light curve fitting stage. The 
resulting posteriors are shown in Figure~\ref{fig:histosg}. 

The n-MAP MCMC stage stopped when $3.75\times10^6$ trials had been accepted,
using 1\% jump sizes for all six free parameters ($e_1$, $\omega_1$, 
$e_2$, $\omega_2$, $e_3$, $\omega_3$). Mixing was checked by evaluating the 
effective lengths of all free parameters from the non-burnt trials, and we found 
the lowest effective length was 1098 for parameter $e_1$. In order to keep only 
$1.0\times10^6$ points in the figures, we evenly down-sampled the chain to 
produce the figures. Table~\ref{tab:diags_g} summarizes the MCMC diagnostics.

\begin{figure*}
\begin{center}
\includegraphics[width=16.8 cm]{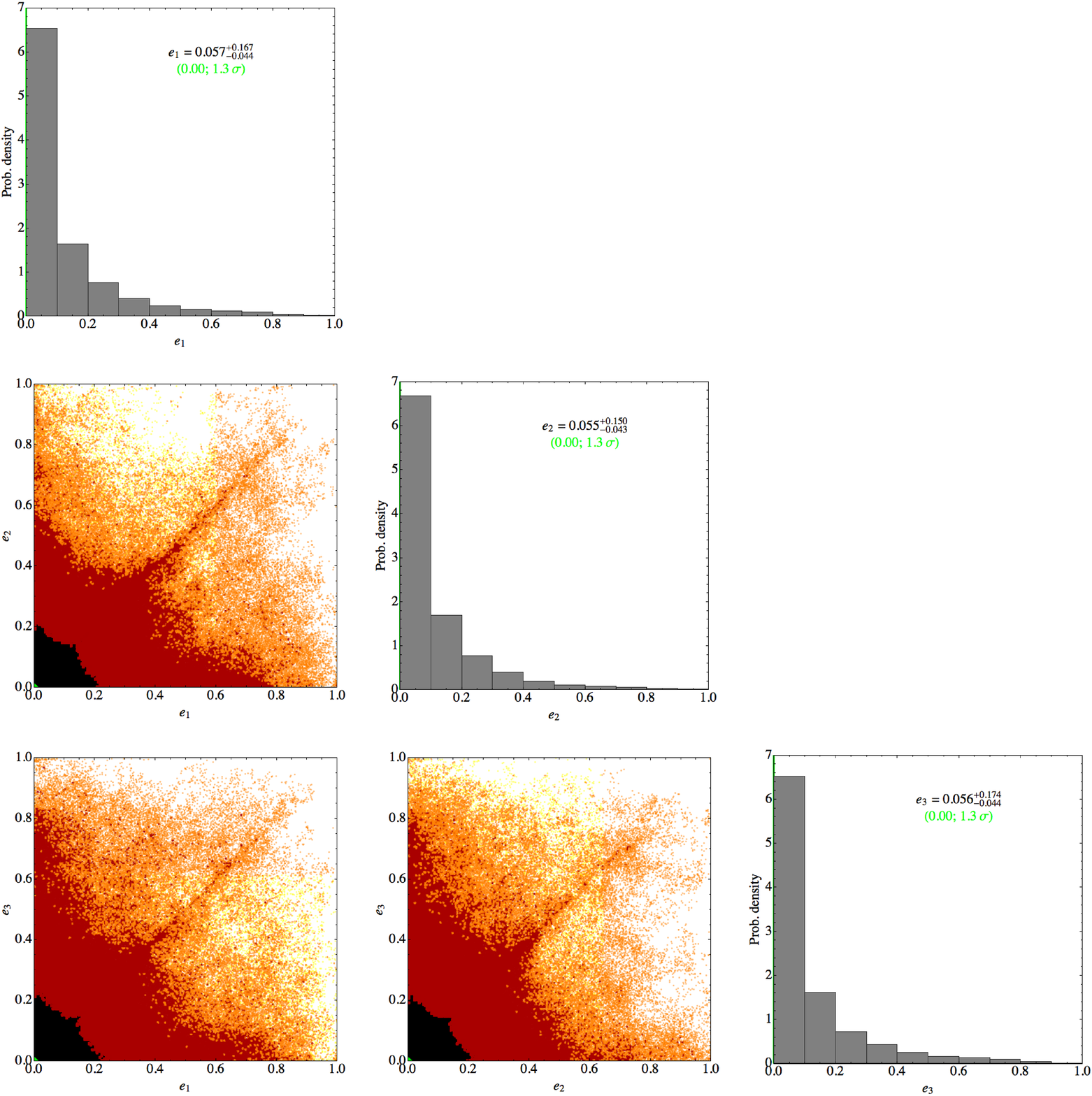}
\caption{\emph{
Results of the numerical MAP fits for KOI-S0G, using uniform priors in $e_k$
and $\omega_k$. In all figures, the green lines and diamonds mark the truth.
Black denotes 0-1\,$\sigma$, red denotes 1-2\,$\sigma$, orange 
denotes 2-3\,$\sigma$ and yellow denotes $>3$\,$\sigma$. Blue lines mark
the median of the lower limits found using analytic MAP (shown in 
Figure~\ref{fig:histosg}). White regions were
never vistted in any MCMC realizations and thus are highly improbable.}} 
\label{fig:nmapg}
\end{center}
\end{figure*}

Since all values were selected to be equal, we expect the system to be 
compatible with a triple-circular orbit system. The n-MAP results indeed agree 
with this conclusion, as shown in Figure~\ref{fig:nmapg}. The bulk of the black 
points ($<1$-$\sigma$) land close to circular orbit solutions and the
marginalized posteriors of $e_k$ reflect the strong preference towards a 
low-eccentricity system.

\begin{table*}
\caption{\emph{MCMC diagnostics of fits for KOI-SOG.}} 
\centering 
\begin{tabular}{c c c c c c} 
\hline
Planet & \# of Accepted & Lowest eff. & Parameter w/ & Highest & Parameter w/ \\ [0.5ex] 
 & MCMC trials & length & lowest eff. len. & Geweke diag. & highest Geweke diag. \\ [0.5ex] 
\hline
KOI-S0G.01 & $1\times(1.00\times10^5)$ & N/A & N/A & N/A & N/A \\ 
KOI-S0G.02 & $1\times(1.00\times10^5)$ & N/A & N/A & N/A & N/A \\
KOI-S0G.03 & $1\times(1.00\times10^5)$ & N/A & N/A & N/A & N/A \\
\hline
n-MAP & $1\times(3.25\times10^6)$ & 2978 & $e_1$ & N/A & N/A \\ [1ex]
\hline\hline 
\end{tabular}
\label{tab:diags_g} 
\end{table*}

Both the a-MAP and n-MAP results support the conclusion of a near-circular
triple system. The 90\% upper limits on the eccentricity were found to be 
$e_1<0.31$, $e_2<0.28$ and $e_3<0.33$.

If we had not used n-MAP but assumed uniform priors in $e_k$, the probability
that $e_1<0.3$ (which is a useful rough limit for a habitable world) would be 
30\% and the probability that $e_1>0.3$ would be 70\%. Therefore it would be 
$\sim0.4$ times more likely that the orbit was $e_1<0.3$ than otherwise. The 
same is of course true for $e_2$ and $e_3$. Using n-MAP these odds ratios become 
8.4, 10.4 and 7.8 for $e_1<0.3$, $e_2<0.3$ and $e_3<0.3$ respectively, 
demonstrating the extra information we have gained from using n-MAP.


\subsection{KOI-S0C: A Perfectly Circular Triple-System}
\label{sub:KOI-S0C}

As a final test, we considered a system of three transiting planets, all on
circular orbits, denoted KOI-S0C. This system is identical to KOI-S01 (see
\S\ref{sub:KOI-S01}) except that $e_k = 0$ for all $k$. The system is generated, 
noised, re-fitted and treated with a-MAP and n-MAP with precisely the same 
methodology used on KOI-S01. The MCMC diagnostics are presented in 
Table~\ref{tab:diags_c}, which indicate excellent mixing and convergence in
all cases.

As expected, the correct radii, transit epoch and orbital periods were easily
found in the blind-search. The derived stellar densities, assuming a circular
orbit, were found to be 
$(\rho_{*,1}^{\mathrm{circ}})^{2/3} = 1.255_{-0.011}^{+0.011}$\,g$^{2/3}$\,cm$^{-2}$,
$(\rho_{*,2}^{\mathrm{circ}})^{2/3} = 1.2583_{-0.0058}^{+0.0023}$\,g$^{2/3}$\,cm$^{-2}$ and
$(\rho_{*,3}^{\mathrm{circ}})^{2/3} = 1.252_{-0.074}^{+0.078}$\,g$^{2/3}$\,cm$^{-2}$
(truth is 1.411\,g\,cm$^{-3}$).

\begin{table*}
\caption{\emph{MCMC diagnostics of fits for KOI-SOC.}} 
\centering 
\begin{tabular}{c c c c c c} 
\hline
Planet & \# of Accepted & Lowest eff. & Parameter w/ & Highest & Parameter w/ \\ [0.5ex] 
 & MCMC trials & length & lowest eff. len. & Geweke diag. & highest Geweke diag. \\ [0.5ex] 
\hline
KOI-S0C.01 & $1\times(1.25\times10^5)$ & 7812 & $b$ & 0.0034 & $\tau$ \\ 
KOI-S0C.02 & $1\times(1.25\times10^5)$ & 12499 & $b$ & 0.0017 & $\tau$ \\
KOI-S0C.03 & $10\times(1.25\times10^5)$ & 1083 & $b$ & 0.016 & $\mathrm{OOT}_{-3}$ \\
\hline
n-MAP & $1\times(2.50\times10^6)$ & 2442 & $e_2$ & N/A & N/A \\ [1ex]
\hline\hline 
\end{tabular}
\label{tab:diags_c} 
\end{table*}

The n-MAP MCMC stage stopped when $B=2.5\times10^6$ trials had been accepted,
using 1\% jump sizes for all six free parameters ($e_1$, $\omega_1$, 
$e_2$, $\omega_2$, $e_3$, $\omega_3$). Mixing was checked by evaluating the 
effective lengths of all free parameters from the non-burnt trials, and we found 
the lowest effective length was 1116 for parameter $e_1$. In order to keep only 
$1.0\times10^6$ points in the figures, the chain was again evenly down-sampled.

\begin{figure*}
\begin{center}
\includegraphics[width=16.8 cm]{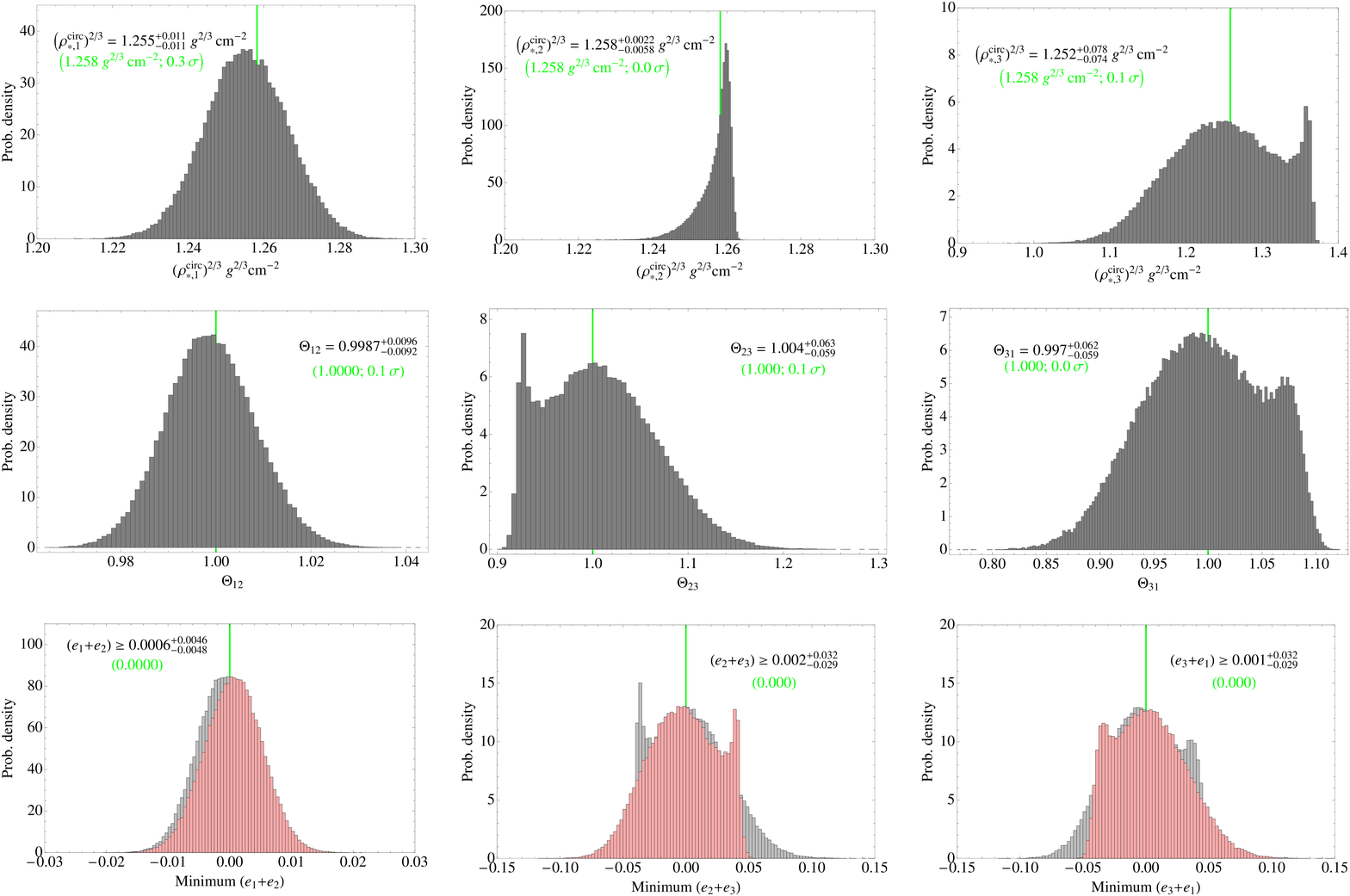}
\caption{\emph{Marginalized posteriors from light curve fits (stage 1) only, 
for KOI-S0C. In all figures, the green line vertical line marks the truth, with 
the numerical value provided in brackets.
\textbf{Row 1:} Posteriors for fitted $(\rho_{*,k}^{\mathrm{circ}})^{2/3}$. 
\textbf{Row 2:} Posteriors for the ratios of the 
$(\rho_{*,k}^{\mathrm{circ}})^{2/3}$ terms.
\textbf{Row 3:} Posteriors for the ratios of the minimum pair-combined
eccentricities, computed using the approximate expressions 
Equation~\ref{eqn:doubleA}\&\ref{eqn:doubleB}.
}} 
\label{fig:histosc}
\end{center}
\end{figure*}

\begin{figure*}
\begin{center}
\includegraphics[width=16.8 cm]{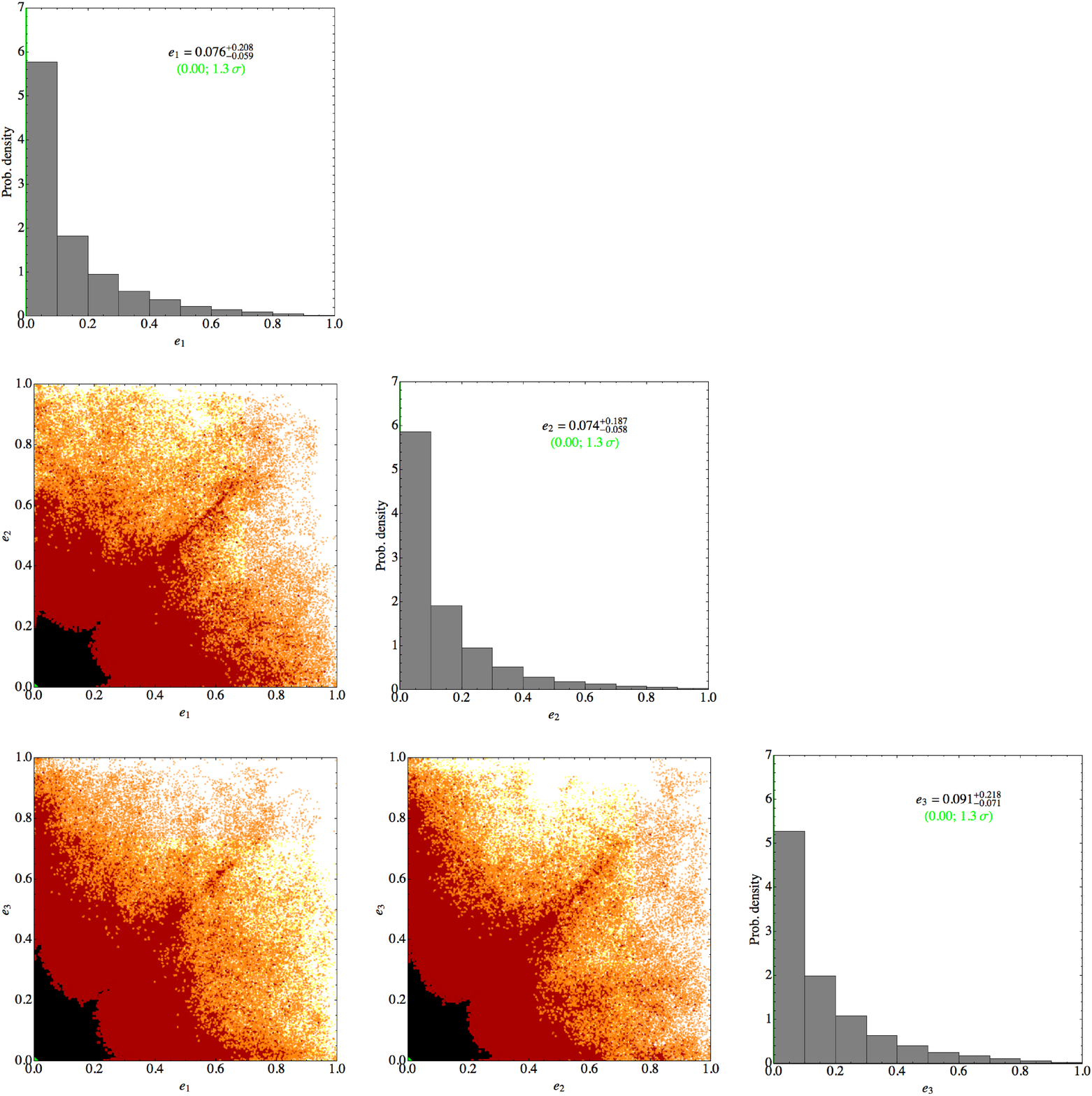}
\caption{\emph{
Results of the numerical MAP fits for KOI-S0C, using uniform priors in $e_k$
and $\omega_k$. In all figures, the green lines and diamonds mark the truth.
Blue lines mark the median of the lower limits found using analytic MAP (shown 
in Figure~\ref{fig:histosc}). Black denotes 0-1\,$\sigma$, red denotes 
1-2\,$\sigma$, orange denotes 2-3\,$\sigma$ and yellow denotes $>3$\,$\sigma$.
White regions were never vistted in any MCMC realizations and thus are highly 
improbable.}} 
\label{fig:nmapc}
\end{center}
\end{figure*}

The results, shown in Figure~\ref{fig:histosc}\&\ref{fig:nmapc}, appear
similar to those of KOI-S0G (\S\ref{sub:KOI-S0G}), which is to be expected
given the fact both cases are consistent with a triply-circular system. The
results again reflect a range of solutions consistent with circular orbits.

As expected, the MAP results suggest that all three-planets are consistent with
a circular orbit. The 90\% upper limits on the eccentricity are $e_1<0.38$, 
$e_2<0.35$ and $e_3<0.41$.

If we had not used n-MAP but assumed uniform priors in $e_k$, the probability
that $e_1<0.3$ (which is a useful rough limit for a habitable world) would be 
30\% and the probability that $e_1>0.3$ would be 70\%. Therfore it would be 
$\sim0.4$ times more likely that the orbit was $e_1<0.3$ than otherwise. The 
same is of course true for $e_2$ and $e_3$. Using n-MAP these odds ratios become 
5.8, 6.8 and 5.0 for $e_1<0.3$, $e_2<0.3$ and $e_3<0.3$ respectively, 
demonstrating the extra information we have gained from using n-MAP.

\bsp

\label{lastpage}

\end{document}